# Multiple Flat Bands and Topological Hofstadter Butterfly in Twisted Bilayer Graphene Close to the Second Magic Angle

Xiaobo Lu[1]†*, Biao Lian[2]†, Gaurav Chaudhary[3]†, Benjamin A. Piot[4], Giulio Romagnoli[5], Kenji Watanabe[6], Takashi Taniguchi[6], Martino Poggio[5], Allan H. MacDonald[3], B. Andrei Bernevig[7] and Dmitri K. Efetov[1]*

[1]ICFO - Institut de Ciencies Fotoniques, The Barcelona Institute of Science and Technology, Castelldefels, Barcelona 08860, Spain.

[2]Princeton Center for Theoretical Science, Princeton University, Princeton, New Jersey 08544, USA.

[3]Department of Physics, University of Texas at Austin, Austin, Texas 78712, USA.

[4]Université Grenoble Alpes, Laboratoire National des Champs Magnétiques Intenses, UPS-INSA-EMFL-CNRS-LNCMI, 38000 Grenoble, France

[5]Department of Physics, University of Basel, Basel 4056, Switzerland

[6]National Institute for Materials Science, 1-1 Namiki, Tsukuba 305-0044, Japan

[7]Department of Physics, Princeton University, Princeton, New Jersey 08544, USA

*Correspondence to: xiaolu@phys.ethz.ch or dmitri.efetov@icfo.eu; †These authors contributed equally to this work.

**Moiré superlattices in two-dimensional (2D) van der Waals (vdW) heterostructures provide an efficient way to engineer electron band properties. The recent discovery of exotic quantum phases and their interplay in twisted bilayer graphene (tBLG) has built this moiré system one of the most renowned condensed matter platforms (*1-10*). So far the studies of tBLG has been mostly focused on the lowest two flat moiré bands at the first magic angle $\theta_{m1}$ ~ 1.1°, leaving high-order moiré bands and magic angles largely unexplored. Here we report an observation of multiple well-isolated flat moiré bands in tBLG close to the second magic angle $\theta_{m2}$ ~ 0.5°, which cannot be explained without considering electron-election interactions. With high magnetic field magneto-transport measurements, we further reveal a qualitatively new, energetically unbound Hofstadter butterfly spectrum in which continuously extended quantized Landau level gaps cross all trivial band-gaps. The connected Hofstadter butterfly strongly evidences the topologically nontrivial textures of the multiple moiré bands. Overall, our work provides a new perspective for understanding the quantum phases in tBLG and the fractal Hofstadter spectra of multiple topological bands.**

Twisted bilayer graphene (tBLG) has emerged as a rich platform to study strong correlations (*1*), superconductivity (*2-8*), magnetism (*9*) and band topology (*10*). Stacking two graphene sheets with a slight rotation by an angle $\theta$ creates a moiré super potential. The renormalized bands have $C_{6z}$, $C_{2x}$ rotational symmetries, and time reversal symmetry T. This leads to decoupled valleys (*11*) and the formation of a mini moiré Brillouin zone (Fig. 1A) with two Dirac cones with identical helicity (Fig. 1B), which makes the resulting two lowest moiré bands at each valley topologically different from the two π bands in monolayer graphene in which the total helicity vanishes. They carry a nonzero Dirac helicity $\eta = \pm 2$ (Fig. 1B), which is protected by $C_{2z}T$ symmetry. These attributes prohibit the construction of local symmetric Wannier functions, so defining the nontrivial topology of tBLG (12-15).

The magic angles in tBLG are a series of well-defined twist-angles for which the moiré bands are predicted to become ultra-flat (11). While the tBLG at $\theta_{m1} \sim 1.1°$ is already extensively studied, tBLG at well-defined smaller magic angles (i.e. $\theta_{m2} \sim 0.5°$), in which inhomogeneity of moiré patterns is much more sensitive to twist angle fluctuations (16), has so far rarely been experimentally investigated. Since tBLG at $\theta_{m2}$ offers an exceedingly large moiré unit cell, it gives rise to a multitude of closely packed flat moiré bands. Crucially, the moiré wavelength for tBLG at $\theta_{m2}$ of $\lambda \sim 30$nm is much larger than for $\theta_{m1}$ tBLG or for graphene/hBN moiré superlattices, and hence inducing one magnetic flux per unit cell, required to obtain a Hofstadter butterfly, can be easily achieved at much lower fields ($\sim 6$ times) (17-19). Together these characteristics make tBLG at $\theta_{m2}$ an ideal platform to explore multiple moiré bands and their Hofstadter butterfly spectra which are rendered with interactions and novel band topology.

In this article, we present magneto-transport measurements of tBLG devices, with $\theta$ close to the predicted second magic angle $\theta_{m2} \sim 0.5°$. Our tBLG device is encapsulated with crystallographically non-aligned insulating layers of hexagonal boron nitride (hBN) and its carrier concentration $n$ is capacitively controlled by an underlying graphite layer (Fig. 1D), and calibrated with quantum oscillations in out-of-plane magnetic field $B_\perp$ (SM Methods). Fig. 1F-G show the evolution of the longitudinal resistance $R_{xx}$ and the Hall density $n_H$ as a function of $n$. $R_{xx}$ exhibits peaks and $n_H$ sign changes at equally spaced, integer multiples of $n_s$, $n = sn_s$, and mark the transitions between the individual bands. The resistance peaks are strongly enhanced by a small $B_\perp$-field (Fig. 1E) and develop thermally activated, gapped behavior (SM Fig. S2-4), but are insensitive to parallel $B$-field (Fig. S5). This suggest an orbital origin and agrees with the fact that small $B_\perp$-fields can gap out the $C_{2z}T$ protected Dirac nodes or increase a small band gap.

The non-interacting band structure of tBLG around $\theta_{m2}$ contains eight strongly convoluted 4-fold spin-valley degenerate low-energy flat-bands (Fig. S10B). To account for interaction effects in the flat-bands, we perform Hartree-Fock calculations including Coulomb interactions with random seeds, which allow for spontaneous symmetry breaking (SM Sec. 5). These show that Coulomb interactions spontaneously break $C_{3z}$, while preserving the $C_{2z}T$ symmetry. As a direct consequence, the moiré bands separate above and below the Fermi energy $E_F$, and become almost non-overlapping except for point-like connections.

This is seen in Fig. 1C (for $E_F$ at the charge neutrality point (CNP), $s = 0$) and in Fig. S10A (for $E_F$ at integer band fillings $s = 1, 2, 3, 4$), which show the Hartree-Fock moiré bands in one valley of $\theta = 0.45°$ tBLG. Here, each separate band remains 4-fold degenerate and can intake $n_s = 4/\Omega_m$ carriers, where $\Omega_m$ is the area of the moiré unit cell. The carrier density can be expressed by integer multiples of band fillings $sn_s$, which defines the integer band filling factor $s \in \mathbb{Z}$. For $s = 0, \pm 1, \pm 3, \pm 4$, the band structure is semi-metallic with two Dirac points per spin per valley at $E_F$, while for filling $s = \pm 2$, a small indirect gap occurs. In particular, we find that the helical Dirac nodes at the CNP, which are responsible for the nontrivial topology, are quite robust for both interacting and non-interacting calculations, although they may deviate from high symmetry points due to $C_{3z}$ symmetry breaking.

We compare the Hartree-Fock spectrum with transport measurements of a tBLG device with $\theta \sim 0.45°$, and find good agreement with its non-overlapping bands picture. The experimentally resolved peaks in $R_{xx}$ (equally spaced at $n = sn_s$) and the sharp interband transitions in $n_H$ indicate that the bands are non-overlapping, either semi-metallic or separated by small gaps (SM Fig. S2A), in good agreement with the Hartree-Fock calculations. Since the studied devices have $\theta$ close to the predicted second magic angle $\theta_{m2} \sim 0.5°$, these findings support the existence of flat bands and strong electronic interactions. However, we do not yet observe correlation driven insulating states as in the case of tBLG at $\theta_{m1}$. This may be due to suppression of the on-site Coulomb energy for a larger moiré wavelength ($\sim 10\,\text{meV}$ for dielectric constant $\epsilon \sim 5$), the larger noninteracting bandwidths ($\sim 10\text{meV}$), and the close energetic proximity of the flat-bands.

When the $B_\perp$-field is increased further (Fig. 2A), Landau levels (LL) develop within the individual bands, and form gaps with nonzero Chern numbers ($C = 4, 8, 12 \ldots$ due to spin/valley degeneracy), which manifest themselves as dips in $R_{xx}$ (blue). These dips trace back to the band-edges at integer fillings $n/n_s = s$ at zero $B_\perp = 0T$ and follow a well-defined slope $dn/dB_\perp = Ce/h$ in the $n$-$B_\perp$ phase space, where $e$ is the electron charge, $h$ is Planck's constant. Above $B_\perp > 8T$ robust quantum Hall states develop, with $R_{xx} \sim 0\Omega$, and with quantized plateaus in the Hall resistance of $R_{xy} \sim h/Ce^2$ (Fig. 2C). In contrast, all trivial band-gaps have $C = 0$ and emerge from $B_\perp = 0T$ without a slope in $dn/dB_\perp = 0$. These gaps are identified by $R_{xx}$ peaks (red) that follow straight vertical lines in the $n$-$B_\perp$ phase space. For clarity, we highlight the most pronounced gaps from Fig. 2A in Fig. 2B.

These findings indicate the formation of a fractal Hofstadter spectrum. The trajectories of the trivial band-gaps and the topological LL gaps have a dense set of intersections at which only one of the gaps survives. The intersections occur at well-defined magnetic fields $B_\perp \sim 2.25T$ ($\phi \sim \phi_0/2$), $B_\perp \sim 4.5T$ ($\phi \sim \phi_0$) and $B_\perp \sim 9T$ ($\phi \sim 2\phi_0$), where $\phi = B_\perp \Omega_m$ is the magnetic flux per moiré unit cell, and $\phi_0 = h/e$ is the magnetic flux quantum. The LL gaps in the $n$-$B_\perp$ diagram occur along lines that are described by the Diophantine relation, $n/n_s = C\phi/4\phi_0 + s$. This is further confirmed by $R_{xy}$ versus $n$ measurements in the quantum Hall regime at $B_\perp = 15T$ (Fig. 2C), which show a non-monotonic evolution of the plateaus, consistent with the development of Hofstadter gaps within different moiré bands.

We denote the features of the Hofstadter gaps by $(C, s)$, which have Chern number $C$ and emerge from the band edges at filling $n = sn_s$. The fan diagram is dominated by a series of zero field single particle band-gaps $(0, s)$, and by Hofstadter gaps $(\pm 4, s)$, $(\pm 8, s)$, which are strikingly not confined within the moiré band from which they emerge (at $B_\perp = 0$), but continuously extend into the higher lying moiré bands at high $B_\perp$, and cross all $(0, s)$ band-gaps. For example, the $(\pm 4, 0)$ gaps that are emerging from the CNP (blue diagonal lines), extend indomitably through several higher moiré bands, interrupting all band gaps $(0, s)$ at carrier densities $n = \pm sn_s$ (red vertical lines). The observed Hofstadter spectrum is qualitatively distinct from that of topologically trivial bands.

We highlight the differences by comparing typical Hofstadter spectra of topologically trivial and non-trivial bands (Fig. 3A-B). For separated topologically trivial bands, the Hofstadter spectrum is confined within the energy bandwidth of each band. As $B_\perp$ increases, LLs from the band-top and band-bottom move towards the middle of the band and are annihilated around its center. This evolution of LLs in $B_\perp$ field defines an energetically bounded Hofstadter butterfly, where the LL spectra in one band are not connected to those of other bands (while one can devise trivial band structures with a connected spectrum, they are not generic). This is in stark contrast to the Hofstadter butterfly which is formed from topologically non-trivial bands (Fig. 3B). In all generality, the Hofstadter spectrum of a topological band is not energetically confined to the bandwidth of the band and can propagate until it connects to the Hofstadter spectrum of another band, which trivializes the total band topology (14, 20).

Such a topologically non-trivial, unbound and connected Hofstadter butterfly spectrum is clearly present in the demonstrated $\theta = 0.45°$ tBLG device, as well as in several other devices with $\theta = 0.3°$ - $0.5°$ (SM Sec. 3). Fig. 3C displays a zoom-in of the dashed region in Fig. 2A, and highlights the interplay between the trivial band-gaps $(0, s)$ (red) and the Hofstadter gaps $(\pm 4, 0)$ (blue), where $(0, \pm 2)$ are clearly interrupted by $(\pm 4, 0)$ at $B_\perp \sim 9\text{T}$ ($\phi/\phi_0 = 2$). Similarly $(0, \pm 1)$ are interrupted by $(\pm 4, 0)$ at $B_\perp \sim 4.5\text{T}$ ($\phi/\phi_0 = 1$). These interruptions indicate the closing of the $(0, \pm s)$ band-gaps between two neighboring bands, when their fillings coincides with that of the $(\pm 4, 0)$ LL gaps. This connects the Hofstadter spectra of the lowest two moiré bands with the spectra of all higher moiré bands, which is a direct evidence of the nontrivial topology of tBLG, and is in agreement with the theoretical predictions in Ref.(14, 20).

The experimentally obtained unbounded and connected Hofstadter spectrum is in good agreement with the theoretically calculated Hofstadter spectrum for low $n$ ($|n/n_s| < 3$) (Fig.3D and MS Fig. S12B). It is calculated with a tBLG continuum model at $\theta = 0.45°$, which includes broken $C_{3z}$ symmetry due to strain or Hartree-Fock mean fields (SM Sec. 6), to realistically mimic a strain reconstructed tBLG device. We find that the calculated $(\pm 4, 0)$ gaps continuously extend to the higher bands as expected. Moreover, experimentally we observe reemerged $(0,0)$ and $(0, \pm 1)$ gaps at $\phi > \phi_0$, and $(0, \pm 2)$ gaps in the range $\phi_0 < \phi < 2\phi_0$, which match the calculations only if $C_{3z}$ breaking is considered (SI Fig. S12). However, the observed Hofstadter spectra at high doping densities ($|n/n_s| > 3$) show deviations from the theoretical calculations. Theoretical calculations predict a series of $(4, s)$ and $(-4, -s)$ gaps ($s = 1,2,3, \cdots$) at $\phi > \phi_0$ in addition to the $(\pm 4, 0)$ gaps, while experimentally this region is dominated by the $(8,1)$ and $(-8, -1)$ gaps, and only a $(4,1)$ gap occurs at $\phi > 3\phi_0$. This discrepancy may have its origins in the non-negligible role of interactions in the flat-bands, which may considerably affect the Hofstadter spectrum of tBLG at large doping densities. In SM Sec. 6, we show that the inclusion of an electron doping induced Hartree term in the Hofstadter spectrum calculation suppresses the $(4, s)$ gaps, which suggests that interactions may be responsible for their experimental absence.

One possible origin of the nontrivial unbound Hofstadter butterfly is the single-particle fragile topology hosted by the lowest two moiré bands of each spin and valley in tBLG (theoretically predicted and characterized by the $C_{2z}T$ protected nonzero Dirac helicity $\eta = \pm 2$ at the CNP) (12-14). This fragile topology is further enhanced into a stable topology if the particle-hole symmetry of tBLG is preserved, which is robust against adding trivial particle-hole symmetric pairs of bands (15). It is theoretically shown that the Hofstadter spectrum of the lowest two topological moiré bands of tBLG is always connected with the spectrum of the higher moiré bands at sufficiently strong magnetic fields (up to infinity) (14) (although not universally protected by $C_{2z}T$ (20), see

SM Sec. 4). For a wide range of twist angles (including those studied in this paper), the $(\pm 4, 0)$ LL gaps are predicted to extend from the lowest two moiré bands to all the higher bands, forcing their Hofstadter spectra to be connected at $\phi/\phi_0 = 1$ (*14*). Accordingly, the band gap between two bands with connected Hofstadter spectra will close when the connection happens. We note that the flat moiré bands in tBLG can also develop correlated states of different stable topologies breaking $C_{2z}T$, for instance, Chern insulators at the first magic angle, which is driven by interactions and stabilized by finite magnetic fields (*21-26*). At this point our results cannot radically rule out the possibility that either type of topology, $C_{2z}T$-preserved fragile topology or $C_{2z}T$-broken stable topology, is responsible for the connected Hofstadter spectra. The absence of correlated states in our experiment and the unbroken $C_{2z}T$ symmetry in our Hartree-Fock calculations, however, suggest the former is more likely.

In summary, we have reported a systematic magneto-transport study of tBLG close to $\theta_{m2}$. Our results show a) multiple well-isolated flat moiré bands in tBLG at $\theta_{m2}$, which cannot be explained without considering interactions; b) that tBLG is a highly tunable platform to study the Hofstadter butterfly, where the band topology can manifest itself qualitatively (trivial or nontrivial); c) that interaction effects play a clear role in tBLG Hofstadter spectra that merit further exploration.

**Acknowledgments:** We are grateful for fruitful discussions with Francisco Guinea. D.K.E. acknowledges support from the Ministry of Economy and Competitiveness of Spain through the "Severo Ochoa" program for Centres of Excellence in R&D (SE5-0522), Fundació Privada Cellex, Fundació Privada Mir-Puig, the Generalitat de Catalunya through the CERCA program, the H2020 Programme under grant agreement n° 820378, Project: 2D•SIPC and the La Caixa Foundation. A.H.M. and G.C. acknowledge support from DOE grant DE-FG02-02ER45958 and Welch foundation grant TBF1473. B. L. acknowledge the support from the Princeton Center for Theoretical Science, Princeton University. B.A.B. was supported by the Department of Energy Grant No. de-sc0016239, the Schmidt Fund for Innovative Research, Simons Investigator Grant No. 404513, and the Packard Foundation. Further support was provided by the National Science Foundation EAGER Grant No. DMR 1643312, NSF-MRSEC DMR-1420541, BSF Israel US foundation No. 2018226, ONR No. N00014-20-1-2303, and Princeton Global Network Funds. M. P. thanks the SNF Sinergia network "Nanoskyrmionics" (grant CRSII5-171003).

**Author contributions:** D.K.E. and X.L. conceived and designed the experiments; X.L. performed device fabrication and transport measurements with support from G.R., B.A., M.A. and M.P.; X.L. B.L., G.C., A.H.M., B.A.B. and D.K.E. analyzed the data; B.L., G.C., A.H.M. and B.A.B. performed the theoretical modeling of the data; K.W. and T.T. provided hBN crystals; X.L., B.L., G.C., A.H.M., B.A.B. and D.K.E. wrote the paper with input from all authors.

**Data and materials availability:** The data that support the findings of this study are available from the corresponding author upon reasonable and well-motivated request.

**Supplementary Materials:**

Materials and Methods

Supplementary Text

Figures S1-S14

Captions for Figures S1 to S14

References and notes (*27-33*)

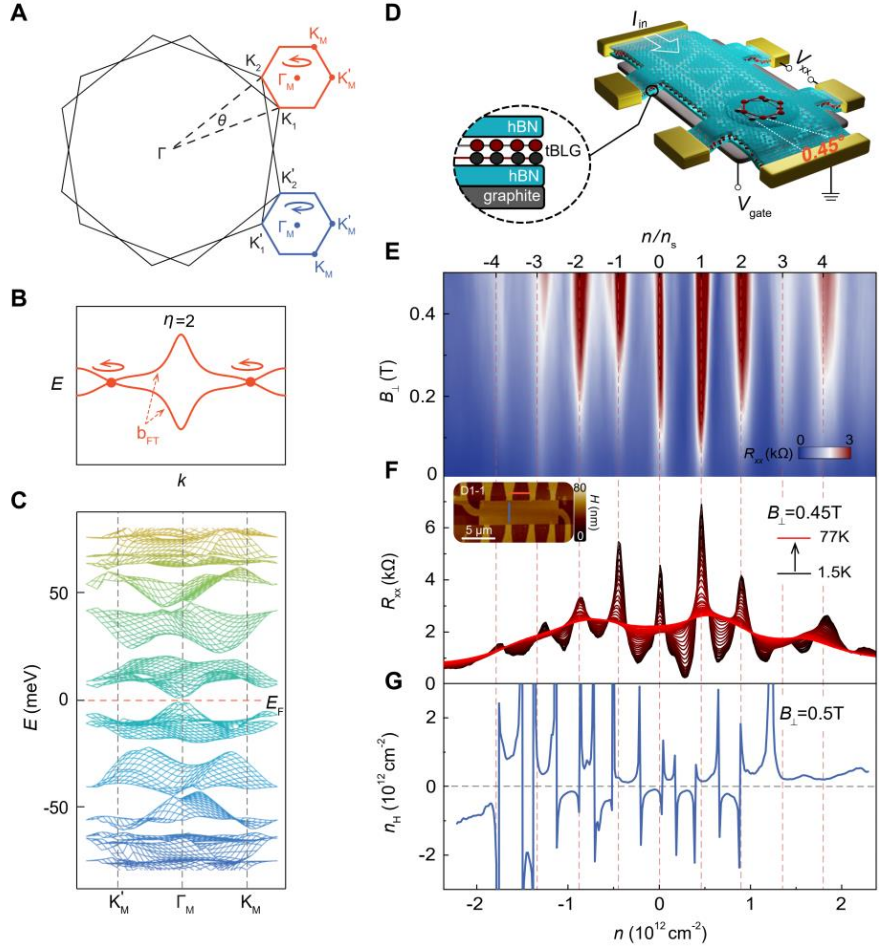

**Fig. 1. Multiple moiré bands with non-zero helicity in $\theta = 0.45°$ tBLG.** (**A**) Mini moiré Brillouin zones are defined by two Dirac cones with a total Dirac helicity $\eta = \pm2$. (**B**) Non-zero total Dirac helicity in the lowest two moiré bands inside of one valley. (**C**) Hartree-Fock band structure calculations of 0.45° tBLG with Coulomb interaction show multiple moiré bands, which are separated at the Fermi energy $E_F$. (**D**) Schematic of the hBN encapsulated $\theta = 0.45°$ tBLG device with a graphite bottom gate. (**E**) Color plot of the longitudinal resistance $R_{xx}$ vs. carrier density $n$ and out-of-plane magnetic field $B_\perp$. (**F**) $R_{xx}$ vs. $n$ at different temperatures and $B_\perp = 450$ mT. Insert displays the AFM image of a typical tBLG device. (**G**) Hall carrier density $n_H = -B_\perp/(eR_{xy})$ as a function of gate induced carrier density $n$. measured at $B_\perp = 500$ mT.

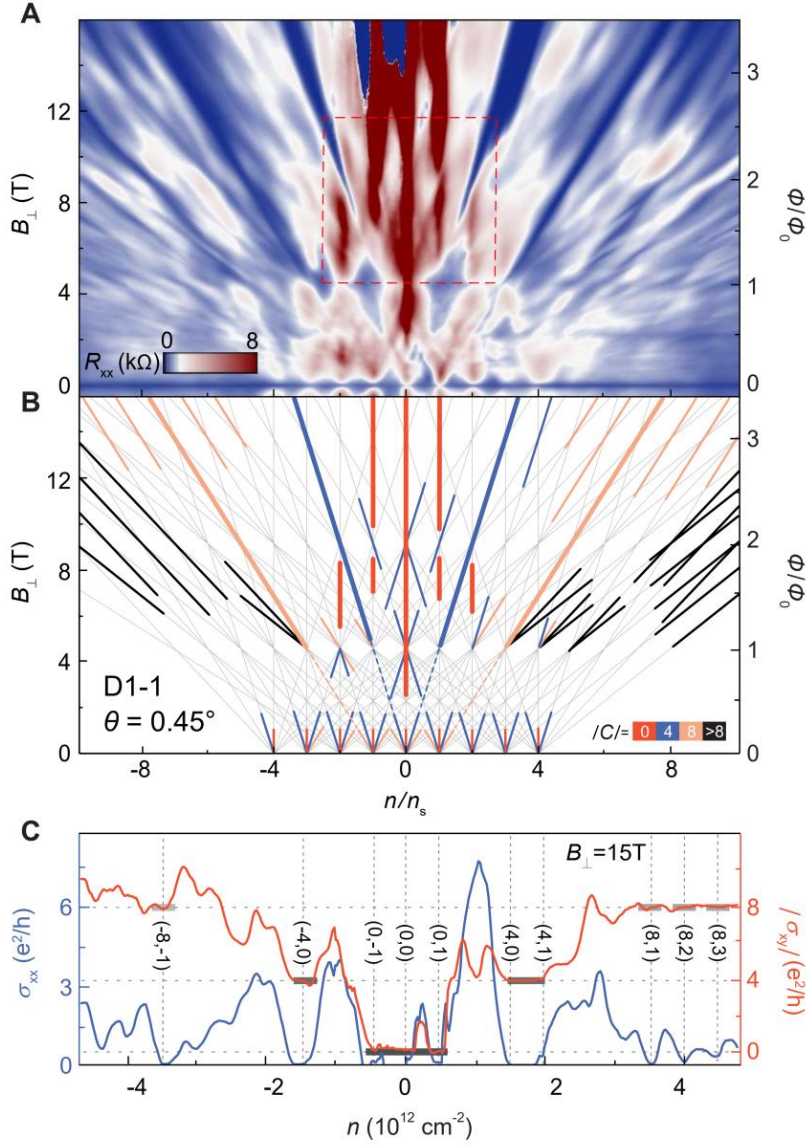

**Fig. 2. Unbounded and connected Hofstadter butterfly spectrum.** (**A**) Hofstadter spectrum of the $\theta = 0.45°$ tBLG device, revealed by the color plot of $R_{xx}$ vs. carrier density $n$ and $B_\perp$. (**B**) Schematic identifying visible LL gaps from (**A**). Vertical solid lines indicate single particle band-gaps $(0, s)$ between the moiré bands, and diagonal solid lines indicate Hofstadter LL gaps $(C, s)$, which emerge from the band edges. (**C**) Longitudinal conductance $\sigma_{xx}$ and Hall conductance $|\sigma_{xy}|$ vs. $n$, measured at $B_\perp = 15T$, show robust quantum Hall states with a non-monotonic evolution with $n$, as expected from the Diophantine relation.

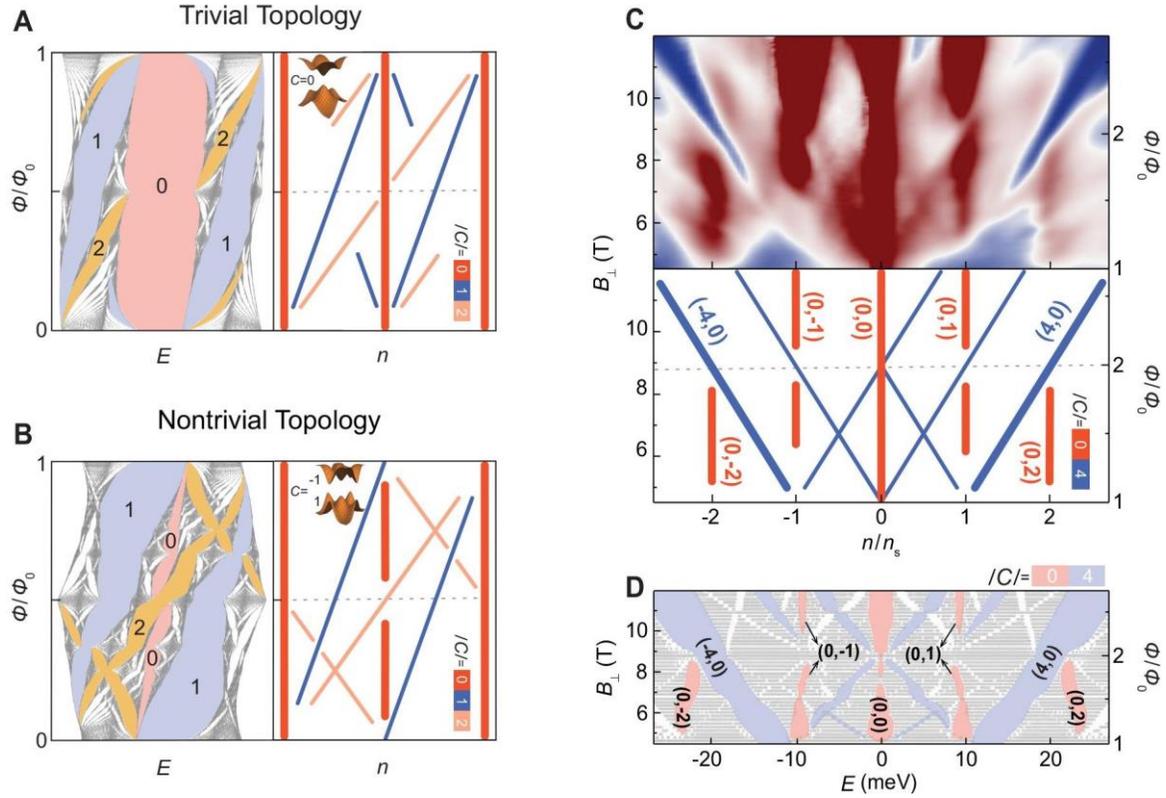

**Fig. 3. Hofstadter butterflies with nontrivial topology.** (**A-B**) Examples of Hofstadter spectra (left) and corresponding gap schematics (right) of two topologically trivial (**A**) and non-trivial $C = \pm1$ (**B**) bands. The topological Hofstadter spectrum shows distinct features, as its bands are connected at $\phi/\phi_0 = 1$, and the $C = 0$ band gaps are closed. Both (**A**) and (**B**) are calculated for tight-binding models (see SM Sec. 9). (**C**) Zoom-in on the Hofstadter spectrum in Fig. 2A-B. The $(\pm4, 0)$ LL gaps continuously extend into the higher lying moiré bands and interrupt the $(0, \pm1)$ and $(0, \pm2)$ band-gaps, showing the main signatures of the topological Hofstadter spectrum. (**D**) Calculated Hofstadter spectrum of $\theta = 0.45°$ tBLG shows good agreement with experiment (**C**).

# Supplementary Materials for

## Multiple Flat Bands and Topological Hofstadter Butterfly in Twisted Bilayer Graphene Close to the Second Magic Angle

Xiaobo Lu[1]†*, Biao Lian[2]†, Gaurav Chaudhary[3]†, Benjamin A. Piot[4], Giulio Romagnoli[5], Kenji Watanabe[6], Takashi Taniguchi[6], Martino Poggio[5], Allan H. MacDonald[3], B Andrei Bernevig[7], Dmitri K. Efetov[1]*.

†These authors contributed equally to this work.

Correspondence to: xiaolu@phys.ethz.ch or dmitri.efetov@icfo.eu

**This PDF file includes:**

> Materials and Methods
> Supplementary Text
> Figs. S1 to S14
> Captions for Figures S1 to S14
> References and Notes:

**Materials and Methods**

Device fabrication

The hBN/tBLG/hBN/graphite stacks were exfoliated and assembled using a van der Waals assembly technique. Monolayer graphene, thin graphite and hBN flakes (~10 nm thick) were first exfoliated on SiO₂ (~300 nm)/Si substrate, followed by the "tear and stack" technique with a polycarbonate (PC)/polydimethylsiloxane (PDMS) stamp to obtain the final hBN/tBLG/hBN/graphite stack. The separated graphene pieces were rotated manually by the twist angle ~1°. We purposefully chose a larger twist angle during the heterostructure assembly due to the high risk of relaxation of the twist angle to the random lower values. To increase the structural homogeneity, we further carried out a mechanically cleaning process to squeeze the trapped blister out and release the local strain. We did not perform subsequent high temperature annealing to avoid twist angle relaxation. We further patterned the stacks with PMMA resist and CHF₃+O₂ plasma and exposed the edges of graphene, which was subsequently contacted by Cr/Au (5/50 nm) metal leads using electron-beam evaporation (Cr) and thermal evaporation (Au).

Measurement

Transport measurements were carried out in a refrigerator with a base temperature of 1.3K and up to 16T magnetic field. All the data without specific notification was measured at the base temperature. Standard low-frequency lock-in techniques were used to measure the resistance $R_{xx}$ and $R_{xy}$ with an excitation current of ~10 nA at a frequency of 19.111Hz. A global gate voltage (+20 V) through Si/SiO₂ (~300 nm) is applied to reduce the contact resistance by tuning the charge carrier density separately in the device leads.

Twist angle extraction

The twist angle $\theta$ is deduced from the area of the moiré unit $\Omega_m$ given by $\Omega_m = \sqrt{3}a^2/4(1 - \cos\theta)$, where $a$ is the lattice constant of graphene. $\Omega_m$ is extracted with two independent methods.

Method 1: Gate induced carrier density is first calibrated with LLs. The carrier density of LL fanning out from CNP with total Chern number $C$ strictly follows $n = C$eB/h. Here we use LL with index $C = 8$ to calibrate the carrier density. The area of moiré superlattice $\Omega_m = 4/n_s$ where $n_s$ is the carrier density of a completely filled moiré band.

Method 2: The Hofstadter butterfly is also used to extract $\Omega_m$. When the magnetic flux per moiré unit cell given by $\phi = B\Omega_m$ equals to the magnetic flux quantum $\phi_0 = h/e$ ($h$, Planck's constant; e, magnitude of the electron charge), the Hofstadter spectrum will exhibit fractal signature (i.e. $B \sim 2.25$T, $B \sim 4.5$T and $B \sim 9$T corresponding to $\phi = \phi_0/2$, $\phi = \phi_0$ and $\phi = 2\phi_0$, respectively). The moiré area $\Omega_m$ can be described by $\Omega_m = \phi_0/B_0$ ($B_0$ is the magnetic field corresponding to $\phi = \phi_0$) where the straight lines of LL gaps in the $n$-$B$ plot cross each other at integer band fillings.

**Supplementary Text**

**1. Device information**

We have measured devices with four different twist angles [D1-1 (0.45°), D1-2 (0.44°), D2 (0.38°) and D3 (0.34°)]. Fig. S1 shows optical microscope images of corresponding stacks and devices.

**2. Gap opening at low magnetic field**

Fig. S2 shows the temperature dependent $R_{xx}$ vs. $n$ measured at different magnetic field $B_\perp$. At zero magnetic field, all integer filling states exhibit a metallic behavior (Fig. S2A). Interestingly, at finite $B_\perp$, these states start becoming thermally activated, showing an insulating behavior (Fig. S2B-D). As shown in Fig. S3, the gap values are extracted by fitting $R_{xx}\sim\exp(\Delta/2kT)$ temperature activated behavior for all integer filling states. We further demonstrate the evolution of gaps with magnetic field $B_\perp$ (Fig. S4) for these state. All these gaps increase when magnetic field $B_\perp$ becomes large until quantum Hall regime starts to dominate the transport. To understand the magnetic field induced metal-insulator transition in tBLG better, we have further measured the device D1-1 with tilted magnetic field (Fig. S5). By comparing Fig. S5a and Fig. S5B, we have found that the resistivity of all integer filling states is only sensitive to the perpendicular component of magnetic field. Our results indicate that it is orbital effect instead of spin that leads to the gap opening here. This is consistent with the fact that the Dirac nodes at a certain integer filling (if semi-metallic) are protected by a $C_{2z}T$ symmetry which does not act on spin (where T is the spinless time-reversal operator which does not act on spin) (*13*). Since in-plane field only acts on spin, it cannot gap out the Dirac nodes in the semimetallic case. In the case where the system has no Dirac nodes but a small gap, a small $B_\perp$ would produce Landau levels and enlarge the gap, while an in-plane field shifts the spin up and down (pointing in-plane) bands oppositely and thus generically reduces the gap, which is also consistent with our experimental observation

**3. Hofstadter butterfly of tBLG with different angles**

In Fig. S7-8, we demonstrate magneto-transport results of three more devices (D2 0.38°, D3 0.34° and D1-2 0.44°). Similar with the spectrum shown in Fig. 2 in main text, all samples we have measured exhibit robust unbounded and connected features in Hofstadter spectrum. As shown continuous extension of LLs from the CNP to higher moiré bands shows a clear evidence of nontrivial band topology of tBLG, which otherwise would be obstructed. Similar with D1-1 shown in main text, $(4,0)$, $(-4,0)$, $(8,1)$ and $(-8,-1)$ LL gaps are still dominant in the Hofstadter spectrum and crossing several moiré bands, giving an unbounded and connected Hofstadter butterfly in all devices. Some $C=0$ gaps also reappear in magnetic field, i.e. the $(0,0)$ gap starts to appear when the magnetic flux through one moiré unit cell is close to half quantized value $\phi_0/2$, whereas $(0,\pm1)$ gaps reappear when the magnetic flux is above $\phi_0$. We find that $(0,\pm1)$ gaps in all measured samples are interrupted by $(\pm4,0)$ gaps that are emerging from the CNP. Moreover, $(0,\pm2)$ gaps in device D2 and $(0,-2)$ in device D1-2 also reappear above $\phi_0$ and are further interrupted by $(\pm4,0)$ gaps.

**4. Hofstadter butterfly of fragile topology**

As demonstrated in Ref (*14*), the Hofstadter butterfly of the lowest two fragile topological bands of the single-valley tBLG continuum model (see SM Sec. 6 for definition) is always connected with the Hofstadter spectra of the higher bands at certain nonzero magnetic fluxes per moiré unit cell, which can be understood as the fingerprint of the fragile topology (this is specifically true for the tBLG continuum model here; see the last paragraph of this section for a brief discussion of generic $C_{2z}T$ fragile topology models). For large enough twist angles ($> 2°$), the perturbation analysis in Ref ($14$) shows that the Dirac helicity $\eta = 2$ at each graphene valley (which gives the fragile topology of the lowest two tBLG moiré bands) leads to two Chern gaps ($\pm 4, 0$) (counting the 4-fold spin-valley degeneracy) extending from the lowest two moiré bands to all the higher bands, which interrupts all the moiré band gaps ($0, s$) at nonzero integer fillings $s = n/n_s$. As a result, the Hofstadter butterflies of all the moiré bands are connected together at nonzero magnetic fields. Numerical calculations in Ref ($14$) show that in a vast range of twist angle and relaxation parameter (see SM Sec. 6 for definition) including the parameters for the tBLG samples in our experiment, the Hofstadter spectrum of the lowest two moiré bands is connected with the spectrum of the higher bands at nonzero magnetic fluxes no larger than $\phi/\phi_0 = 1$.

Here we also show in Fig. S9 (replotted from Ref ($14$)) the calculated Hofstadter butterfly for the ten-band tight-binding model for tBLG (single spin and single valley) proposed by Ref ($12$), which faithfully characterize the fragile topology of the tBLG continuum model. In Fig. S9, we have multiplied all the Chern numbers of the Hofstadter gaps of the tight-binding model by 4, accounting for the 4-fold spin-valley degeneracy. One can clearly see that the Hofstadter spectrum of the lowest two bands is connected with the higher Hofstadter spectra at $\phi = \phi_0$, and the first moiré band gap ($0, \pm 1$) are interrupted by the Chern gaps ($\pm 4, 0$) from the CNP which extend all the way to higher bands.

We note that depending on twist angle and the relaxation parameter, the lowest two moiré bands are not always gapped from the higher bands at zero magnetic field (in the absence of interactions). When there is no gap from the higher bands at zero magnetic field (for example, in the range of angles we study in this paper), the fragile topology of the lowest two moiré bands is in principle ill-defined. However, it is shown in previous work ($13$) that in all the tBLG parameter (twist angle) range where the lowest two moiré bands are gapped from the higher bands, the lowest two moiré bands are fragile topological. This shows that the gap closing and reopening between the lowest two moiré bands and the higher moiré bands in the entire tBLG parameter range do not alter the fragile topology of tBLG, and the Dirac helicity $\eta = 2$ of each graphene valley at the CNP remains robust. As a result, for angles where the lowest two moiré bands are gapless with the higher bands, we find the fingerprint of the fragile topology, that the Hofstadter spectra of the lowest two moiré bands and the higher moiré bands are connected at nonzero magnetic field, remains unchanged.

We also note that, in a generic symmetry analysis, a model with two fragile topological bands protected by $C_{2z}T$ symmetry could have their Hofstadter butterfly disconnected with that of the other bands, due to the T symmetry breaking in the zero field model ($20$). However, the disconnection of the Hofstadter butterfly requires a large enough T symmetry breaking; the $C_{2x}$ $C_{2z}T$ symmetry of the graphene model at any magnetic field also allows for Weyl points on the y-momentum axis. In particular, for the TBG continuum model, by both theoretical and numerical analyses, it is shown in Ref ($14$) that the Hofstadter butterfly connection between the lowest two fragile topological bands and the higher bands (in the range of nonzero magnetic fluxes including infinity) is stable for all the twist angles and relaxation parameters $0 \leq u_0 \leq 1$.

## 5. Bandtructure of 0.45° tBLG and Hartree-Fock calculation

We perform a self-consistent Hartree-Fock (HF) calculation for the continuum model to examine the effects of Coulomb interaction between electrons at integer moiré fillings $n = Nn_s$ ($N \in \mathbb{Z}$) at zero magnetic field. We take the Coulomb interaction

$$H_I = \frac{1}{2} \sum_{\alpha, \beta, q, k, k'} \frac{V(q)}{A_{tot}} c^\dagger_{\alpha, k+q} c_{\alpha, k} c^\dagger_{\beta, k'-q} c_{\beta, k'} \, ,$$

where $V(q) = 2\pi e^2 (1 - \delta_{q,0})/\epsilon q$ is the Fourier transform of Coulomb interaction $e^2/\epsilon r$ subtracting the background charge, $A_{tot}$ is the total area of the system. $c_{\alpha, k}$ is the annihilation operator of the Dirac electron of the monolayer graphene at momentum $k$ (the plane wave basis), and $\alpha$ is a shorthand notation for multiple indices: the sublattice $a = A, B$, layer $l = 1,2$, spin $s = \uparrow, \downarrow$ and graphene valley $\eta = K, K'$ indices (see also SM Sec. 6). The electron eigenstates are given by the single-particle mean-field Hartree-Fock Hamiltonian

$$H = H_0 + \Sigma_H + \Sigma_F \, ,$$

where $H_0$ is the free continuum model Hamiltonian (see SM Sec. 6). The Hartree term and Fock terms take the form

$$\Sigma_H = \sum_{\alpha, \beta, q, k, k'} \frac{V(q)}{A_{tot}} \langle c^\dagger_{\alpha, k+q} c_{\alpha, k} \rangle c^\dagger_{\beta, k'-q} c_{\beta, k'} \, ,$$

$$\Sigma_F = - \left[ \sum_{\alpha, \beta, q, k, k'} \frac{V(q)}{A_{tot}} \langle c^\dagger_{\alpha, k+q} c_{\beta, k'} \rangle c^\dagger_{\beta, k'-q} c_{\alpha, k} + h.c. \right] \, ,$$

where $\langle \mathcal{O} \rangle$ stands for the expectation value of operator $\mathcal{O}$ from all the occupied electrons for a given electron filling $n$. We assume there is neither spontaneous translational symmetry breaking nor spontaneous polarization of spins and valleys. In general, translational symmetry breaking will enlarge the unit cell, and spin/valley polarization will break the 4-fold spin-valley degeneracy, both of which may lead to insulating states at non-integer fillings $n/n_s$. Such non-integer filling insulating states are not observed in our experiment (see main text Fig. 1), so we assume translation symmetry and spin-valley degeneracy are unbroken. Thus $\langle c^\dagger_{\alpha, k} c_{\beta, k'} \rangle$ is nonzero only if $k - k' = m_1 g_1 + m_2 g_2$ ($m_1, m_2 \in \mathbb{Z}$) and $\alpha, \beta$ belong to the same spin and valley, where $g_1$ and $g_2$ are the reciprocal vectors of the moiré superlattice.

To perform the self-consistent calculation, we take a moiré momentum lattice cut off of the single-particle Hamiltonian $H_0$ so that it contains 146 bands (i.e., is a $146 \times 146$ matrix) per spin per valley. In the first step, we add to $H_0$ a small Hermitian random matrix $H_{seed}$ as a seed for all possible spontaneous symmetry breakings, and diagonalize $H_0 + H_{seed}$, after which we calculate the expectation values in $\Sigma_H$ and $\Sigma_F$ at the desired fixed filling. Then we iteratively diagonalize the full Hamiltonian $H = H_0 + \Sigma_H + \Sigma_F$ (the initial seed is no longer added after the first step) and calculate $\Sigma_H$ and $\Sigma_F$ at the fixed filling until convergence (until the total Hartree-Fock energy $E_{HF} = \langle H_0 + \frac{1}{2}(\Sigma_H + \Sigma_F) \rangle$ changes less than 0.01meV in one step). In the calculation of Fig. 1C and Fig. S10, we take $\frac{|g_1| V(g_1)}{\sqrt{3}} = 10$meV, relaxation parameter $u_0 = 0.3$ (see SM Sec. 6), and add the random initial seed $H_{seed}$ to allow for spontaneous symmetry breakings.

Our Hartree-Fock calculations for integer fillings $n = Nn_s$ with an initial Hermitian random matrix symmetry breaking seed $H_{seed}$ show that the $C_{3z}$ symmetry is spontaneously broken, but $C_{2z}T$ is preserved. The Hartree-Fock band structures and the density of states for $N = 0,1,2,3,4$ are shown in Fig. S10A and Fig. S11, respectively. For fillings $n/n_s = 0,1,3,4$, we find the band structures are semimetallic with 2 Dirac points at the Fermi energy. For filling $n/n_s = 2$, the band structure has a small indirect gap at the Fermi energy. Lastly, we comment that the realistic tBLG samples may also have external $C_{3z}$ breakings due to uniaxial strains, as revealed by scanning tunneling microscope experiments of tBLG (*27-30*).

## 6. <u>Hofstadter Butterfly Calculation</u>

We calculate the Hofstadter butterfly of the tBLG continuum model, which can be written in real space as

$$H_0 = \int d^2\boldsymbol{r} \sum_{\eta,s} c^\dagger_{\eta,s}(\boldsymbol{r}) \begin{pmatrix} \hbar v \boldsymbol{\sigma}^\eta \cdot (-i\nabla) & T^\eta(\boldsymbol{r}) \\ T^{\eta\dagger}(\boldsymbol{r}) & \hbar v \boldsymbol{\sigma}^\eta \cdot (-i\nabla) \end{pmatrix} c_{\eta,s}(\boldsymbol{r}),$$

where $c_{\eta,s}(\boldsymbol{r}) = \left( c_{A,t,\eta,s}(\boldsymbol{r}), c_{B,t,\eta,s}(\boldsymbol{r}), c_{A,b,\eta,s}(\boldsymbol{r}), c_{B,b,\eta,s}(\boldsymbol{r}) \right)^T$ is the free electron basis ($c_{a,l,\eta,s}(\boldsymbol{r})$ for an electron in sublattice $a$, layer $l = t, b$ (for top and bottom), valley $\eta$ ($= \pm 1$ for K, K') and spin $s$), $\boldsymbol{\sigma}^\eta = (\eta\sigma_x, \sigma_y)$ are the Pauli matrices, $T^\eta(\boldsymbol{r}) = \sum_j^3 T_j^\eta e^{i\eta\boldsymbol{q}_j \cdot \boldsymbol{r}}$, and the 3 matrices $T_j$ are given by

$$T_j^\eta = w \begin{pmatrix} u_0 & e^{i2\pi\eta(j-1)/3} \\ e^{-i2\pi\eta(j-1)/3} & u_0 \end{pmatrix}.$$

The dimensionless parameter $u_0$ characterizes the lattice relaxation. It equals to 1 when there is no relaxation. For $0.45°$, we estimate the relaxation parameter to be $u_0 = 0.3$ (*30, 31*), which we use in all of our calculations. Because of the presence of $C_{2z}$ symmetry and the absence of spin-orbit coupling, the Hofstadter butterflies of all the 2 spins and 2 valleys are identical (excluding the Zeeman energy). The Zeeman energy is negligibly small ($0 < E_z < 1.85\ meV$ for the magnetic field range in the experiment, assuming the $g = 2$) compared to the band energies, which is thus ignored in our Hofstadter butterfly calculations.

In real samples, interlayer uniaxial strain may exist which breaks the $C_{3z}$ symmetry of tBLG (more information shown in SM Sec. 7). In addition, electron-electron interactions may also spontaneously break the $C_{3z}$ symmetry, as our Hartree-Fock calculations show (SM Sec. 5). We find the Hofstadter butterfly with a $C_{3z}$ symmetry breaking agrees better with the experimental data. In particular, adding $C_{3z}$ symmetry breaking allows us to reproduce the reemergence of the $(0,0)$, $(0,\pm1)$ and $(0,\pm2)$ gaps at high $B_\perp$ field ($\phi > \phi_0$), one of the experimental details.

The strain induced breaking of $C_{3z}$ symmetry is shown in SM Sec. 7, where due to strain the moiré Brillouin zone vectors $\boldsymbol{q}_j$ for $j = 1,2,3$ are no longer related to each other via three-fold rotation. The resultant moiré Brillouin zone is an irregular hexagon. Alternatively, the effect of the $C_{3z}$ symmetry breaking can also be incorporated in the interlayer hopping matrix by changing $T_1^\eta$ into $\gamma T_1^\eta$ ($\gamma$ is a dimensionless parameter) while keeping $T_2^\eta$ and $T_3^\eta$. In our Hofstadter butterfly calculations, we use the first approach to properly take strain into account (SM Sec. 7). We have verified that the qualitative features of Hofstadter butterfly are the same in both approaches.

Furthermore, to partially include the effect of Coulomb interactions between electrons, we add the following term to the Hamiltonian $H_0$ as an approximate Hartree term:

$$H_I = U_0 \int d^2\boldsymbol{r} \sum_{\eta,s} \left( \sum_{j=1}^{6} e^{i\boldsymbol{g}_j \cdot \boldsymbol{r}} \right) c^{\dagger}_{\eta,s}(\boldsymbol{r}) \, c_{\eta,s}(\boldsymbol{r}),$$

where $\boldsymbol{g}_j$ are the six smallest reciprocal vectors of the moiré superlattice. Our Hartree-Fock numerical calculations at zero magnetic field show this is the leading term in the Hartree potential. The coefficient $U_0$ depends on the electron filling fraction. For electron (hole) doping, $U_0 > 0$ ($U_0 < 0$).

The Hamiltonian is diagonalized in the Landau level basis of monolayer graphene $|l, N, Y\rangle$, where $N$ is the Landau level index, $Y$ is the guiding center, and we have chosen the Landau gauge. In the Landau level basis, the matrix elements of the interlayer tunneling Hamiltonian $T^{\eta}(\boldsymbol{r})$ involves the evaluation of the following matrix element:

$$\langle t, N, Y | e^{i\boldsymbol{q}_j \cdot \boldsymbol{r}} | b, M, Y' \rangle = \delta_{Y, Y' + q_{jx}\ell^2} e^{\frac{i q_{jy}(Y+Y')}{2}} \chi_{N,M}(\boldsymbol{q}),$$

$$\chi_{N,M}(\boldsymbol{q}_j) = \begin{cases} e^{-\left(\frac{q_j \ell}{2}\right)^2} \dfrac{\sqrt{M!}}{\sqrt{N!}} \left( \dfrac{(-q_{jx} + i q_{jy})\ell}{\sqrt{2}} \right)^{N-M} L_M^{N-M}\left( \dfrac{q_j^2 \ell^2}{2} \right) & N \geq M \\[2ex] e^{-\left(\frac{q_j \ell}{2}\right)^2} \dfrac{\sqrt{N!}}{\sqrt{M!}} \left( \dfrac{(q_{jx} + i q_{jy})\ell}{\sqrt{2}} \right)^{M-N} L_N^{M-N}\left( \dfrac{q_j^2 \ell^2}{2} \right) & M > N. \end{cases}$$

Here $L_M^N$ are the associated Laguerre polynomials. Hence the interlayer tunneling in the Landau level basis couples a guiding center $Y$ from one layer to another guiding center $Y' = Y \pm \Delta_j$, where $\Delta_j = q_{jx}\ell^2$ from the other layer. Here $\ell$ is the magnetic length. For the $C_{3z}$ symmetric unstrained case, as well as for the $C_{3z}$ broken strained case under our choice of strain parameters (which constraints $q_{2x} = -q_{3x}$) (SM Sec. 7 and Fig. S13B), the interlayer term can be interpreted as a 1D lattice of guiding centers with nearest neighbor hopping and the lattice constant $\Delta = |q_{2x}|\ell^2$.

Similarly, the Hartree term $H_I$, couples the guiding center $Y$ with the guiding centers $Y'' = Y \pm \delta_j$, where $\delta_j = g_{jx}\ell^2$ within the same layer. The values taken by $g_{jx}$ are $0, \pm\Delta/\ell^2, \pm 2\Delta/\ell^2$. In this guiding center chain picture, the Hartree term only further introduces next nearest neighbor hopping.

The guiding center chain becomes periodic for the rational flux ratio through the moiré unit cell, i.e. when

$$A_m \ell^2 = 2\pi \frac{p}{q}, \qquad A_m = 3\sqrt{3} |K|^2 (\theta^2 - \nu^2 \epsilon) \ell^2,$$

For positive integers $p$ and $q$. Here $|K| = 4\pi/3a_0$, and $\epsilon, \nu$ are the strain and Poisson ratio [see SM Sec. 7]. The above relation for the area of the strained moiré Brillouin zone is only valid for the specific strain we have chosen for the purpose of numerical efficiency in Hofstadter butterfly calculation. In the limit of no strain one flux through moiré unit cell corresponds to B~25 $\theta^2$ magnetic field in Tesla. In the guiding center chain picture, the number of orbitals on each guiding center site is set by the number of Landau levels kept within a cut-off to obtain convergent result for the Hofstadter spectrum. Overall the dimension of the matrix to be diagonalized is $(4N_c + 2)q$, where the monolayer graphene Landau levels ranging from $-N_c$ to $N_c$ are kept within the cutoff. The value of $N_c$ is set by $\sim 500\ meV$ energy window on either side of Dirac point, with some variation for smoothing. Rest of the diagonalization procedure follows Bistritzer et. Al (*11*).

The gaps in the Hofstadter spectrum have associated topological indices $(\sigma, s)$, where $\sigma$ refers to the Hall conductivity and $s$ is associated with moiré band filling. To calculate these indices, we first notice the gaps in the Hofstadter spectrum and calculate the Landau level fulling $\nu_{LL}$ at those gaps. Next the Landau level fillings are plotted as function of flux ratio. The gap trajectories plotted this way follow simple relation:

$$\nu_{LL} = \sigma + s\frac{p}{q},$$

and the integers $\sigma$ and $s$ are simply the intercept and the slope of these gap trajectories.

In Fig. S12B-C, we compare the Hofstadter butterflies for $U_0 = 0$ (corresponding to no doping), both of which have a 0.4% strain breaking the C$_{3z}$ symmetry (see SM Sec. 7), and $U_0 = 10$meV (corresponding to electron doping), where one can see the $(4, s)$ gaps (with 4-fold degeneracy from spin and valley) are suppressed by $U_0$. This suggest that the absence of $(4, s)$ gaps for $s \geq 2$ (where the sample is electron doped) in the experiment may be due to interaction effects.

## 7. Uniaxial strain and C$_{3z}$ symmetry

In general, both graphene layers are likely to experience strain that may be unavoidable during fabrication process. The signature of strain is observed in the spectroscopy data at the magic angle, where the C$_{3z}$ rotational symmetry broken moiré pattern is observed (*26-29*). Here we outline the procedure to take strain effects into account in our theoretical calculations.

Let the vectors $\boldsymbol{K}_i$, (for $i = 1,2,3$) denote the momentum space position of the three $K$ valley points of a monolayer of graphene, such that:

$$\boldsymbol{K}_1 = \begin{pmatrix} |\boldsymbol{K}| \\ 0 \end{pmatrix}, K_2 = \begin{pmatrix} -\dfrac{|\boldsymbol{K}|}{2} \\ \dfrac{\sqrt{3}}{2}|\boldsymbol{K}| \end{pmatrix}, K_2 = \begin{pmatrix} -\dfrac{|\boldsymbol{K}|}{2} \\ -\dfrac{\sqrt{3}}{2}|\boldsymbol{K}| \end{pmatrix}, \qquad |\boldsymbol{K}| = \frac{4\pi}{\sqrt{3}a_0}.$$

Rotation of a graphene layer by an angle $\theta$ is generated by the rotation matrix

$$R(\theta) = \begin{pmatrix} \cos\theta & -\sin\theta \\ \sin\theta & \cos\theta \end{pmatrix},$$

while application of uniaxial strain $\epsilon$ in a graphene layer along a strain direction at angle $\varphi$ from $x$ -axis is generated by the strain tensor

$$S(\epsilon, \varphi) = R(\varphi)^{-1} \begin{pmatrix} 1 - \epsilon & 0 \\ 0 & 1 + \nu\epsilon \end{pmatrix} R(\varphi),$$

where $\nu = 0.12$ is the Poisson ratio of graphene and the sign of $\epsilon$ dictates compression or stretch.

We consider the tBLG under strain as a two-step process, first, the top and the bottom graphene layers are rotated by angles $-\frac{\theta}{2}$ and $\frac{\theta}{2}$ respectively as shown in the Fig. S13A, then independent uniaxial strain $S(\epsilon_t, \varphi_t)$ and $S(\epsilon_b, \varphi_b)$ are applied to the two layers as shown in the Fig. S13B. The hexagonal moiré Brillouin zone is constructed by taking the difference between the new positions of the three $K$ valley Dirac points of the two layers.

$$\boldsymbol{q}_i = \boldsymbol{K}_i^t - \boldsymbol{K}_i^b = \left[ S(\epsilon_t, \varphi_t) R\left(-\frac{\theta}{2}\right) - S(\epsilon_b, \varphi_b) R\left(\frac{\theta}{2}\right) \right] \boldsymbol{K}_i .$$

In the limit of zero strain, the three high symmetry moiré vectors

$$\boldsymbol{q}_1 = \begin{pmatrix} 0 \\ -2K\sin\frac{\theta}{2} \end{pmatrix}, \boldsymbol{q}_2 = \begin{pmatrix} \sqrt{3}K\sin\frac{\theta}{2} \\ K\sin\frac{\theta}{2} \end{pmatrix}, \boldsymbol{q}_3 = \begin{pmatrix} -\sqrt{3}K\sin\frac{\theta}{2} \\ K\sin\frac{\theta}{2} \end{pmatrix},$$

shown in right of Fig. S13A are related by $C_{3z}$ rotational symmetry. The resultant moiré Brillouin zone is a regular hexagon.

Generally under strain the $C_{3z}$ symmetry is broken and the moiré Brillouin zone is an irregular hexagon as shown in the Fig. S13B. For the numerical efficiency in the Hofstadter butterfly calculations, we consider the form of strain that deforms the hexagonal moiré Brillouin zone in a specific form shown in the Fig. S13C, which still breaks the $C_{3z}$ symmetry. For this specific kind of strain deformation, the strain in the top and bottom layer is generated by $S(\epsilon, \varphi)$ and $S(-\epsilon, \varphi)$ respectively. The strain angle $\varphi$ is determined under the following constraints on the moiré Brillouin zone vectors $\boldsymbol{q}_i$:

$$q_{1x} = 0, \ q_{2x} = -q_{3x},$$
$$q_{2y} = q_{3y} = -\frac{q_{1y}}{2}.$$

We note that this way of breaking $C_{3z}$ makes $\boldsymbol{q}_1$ different from $\boldsymbol{q}_2$ and $\boldsymbol{q}_3$. The effect of such a deformation on the Hofstadter butterfly is similar to the effect of changing $T_1^\eta$ into $\gamma T_1^\eta$ ($\gamma$ is a dimensionless parameter) while keeping $T_2^\eta$ and $T_3^\eta$, as we have checked numerically. Heuristically, this is because enlarging $\boldsymbol{q}_1$ increases the on-site (momentum site) energy (the Dirac fermion term) difference between $\boldsymbol{k}$ and $\boldsymbol{k} + \boldsymbol{q}_1$, which hop with each other by matrix $T_1^\eta$; perturbatively, this is similar to reducing the magnitude of $T_1^\eta$ while maintaining the Dirac kinetic energy difference (which is proportional to $\boldsymbol{q}_1$). At zero magnetic field, both ways of breaking $C_{3z}$ shift the Dirac points away from $K_M$ and $K_M{}'$ points in a similar way.

## 8. Landau fan in the large magnetic field limit

In the large magnetic field limit, both the experiment and the numerical calculation demonstrate that the Landau fan is dominated by $(4, s)$, $(8, s)$, …, where $s$ is an integer. Here we give a heuristic understanding of these gaps from the zero twist-angle limit. We note that at zero twist angle (assuming AB stacking), there are only two low energy graphene bands which are connected by a quadratic Dirac point of helicity $\eta = \pm 2$ at each graphene valley ($\pm$ signs for original graphene valleys K and K', respectively). So if we view the two graphene valleys as decoupled, the Dirac helicity of each AB-stacked graphene valley matches the fragile topology of the tBLG.

The quadratic Dirac point band touching at valley $\eta$ ($\eta = \pm 1$ for K and K') of the zero-twist-angle bilayer graphene can be effectively described by a $k \cdot p$ Hamiltonian $h_{BLG}^\eta(k) = \eta(k_x^2 - k_y^2)\sigma_x + 2k_x k_y \sigma_y$. In a magnetic field, LLs are developed at the quadratic touching bands at each spin and valley, leading to in total 8 degenerate zero mode LLs, and 4-fold degenerate non-zero-mode LLs with energies linear in $B$. As a result, the system with 4-fold spin-valley degeneracy has LL gaps denoted by $(4t, 0)$ with $t = \pm 1, \pm 2, \cdots$, where $(C, s)$ denotes a gap with Chern number $C$ and electron filling $s$ at zero magnetic field ($s = 0$ for CNP). Fig. S14A illustrates such LLs and the LL gaps between them, where gap A has Chern number $C = 4$, and gap G has a Chern number $C = 8$ (counting the 4-fold spin-valley degeneracy). We note that this LL picture is valid because we are considering magnetic fields corresponding to negligibly small magnetic fluxes per original graphene unit cell; hence, we are not considering the Hofstadter butterfly of the bands of untwisted bilayer graphene in the original graphene Brillouin zone. The spectra are therefore well characterized by the LLs of the $k \cdot p$ quadratic Dirac Hamiltonian $h_{BLG}^\eta(k)$ at two graphene valleys, and no Hofstadter physics in the original graphene BZ is expected.

In twisted bilayer graphene, at sufficiently large magnetic field, the Hofstadter spectrum can be thought of as adiabatically deformed from the LLs in the untwisted limit (see Fig. S14A-B), during which the original LL gaps in the untwisted limit remain open above a certain threshold magnetic field. With the two layers twisted relative to each other, a spatially periodic moiré superlattice potential arises, and defines a moiré unit cell (at zero magnetic field) of area $\Omega_m$. Consider some large magnetic flux per moiré unit cell $\phi = p\phi_0$, where $p$ is an integer (so the moiré superlattice translation symmetry is unbroken). Before we turn on the moiré potential, by Streda formula, the number of electron states each LL can accommodate per moiré unit cell is $p$. Therefore, when the moiré potential is turned on, each LL (per spin per valley) in the untwisted limit will split into $p$ subbands. Each subband $j$ ($1 \leq j \leq p$) has $n_j = 1$ electron state per area $\Omega_m$ and may carry some Chern number $\sigma_j$ ($\sum_{j=1}^p \sigma_j = 1$). Fig. S14B gives some illustrative examples of the splitting of the first LL (at one spin one valley) into $p$ subbands at integer fluxes $p = 2,3,4,5$, and the Chern numbers $\sigma_j$ of the subbands at each flux $p\phi_0$ are denoted by the red numbers. We note that here we only require $p$ to be large enough so that the original LL gaps (e.g., gaps A and G in Fig. S14B) are large enough (larger than the typical zero-magnetic-field moiré band widths) and remain open as the moiré potential is turned on. In our case of twist angles around $0.5°$, this requires $p$ greater than 2 or 3.

When the magnetic flux changes, by Streda formula, the number of electron states $n_j$ per area $\Omega_m$ of a subband $j$ at flux $p\phi_0$ satisfies $dn_j/d\phi = \sigma_j/\phi_0$. Therefore, if $\sigma_j \neq 0$, a subband $j$ at flux $p\phi_0$ cannot uniquely deform into a subband $1 \leq j' \leq p + 1$ at flux $(p + 1)\phi_0$, since $n_j = n_{j'} = 1$. If a subband at flux $p\phi_0$ carry a negative Chern number $\sigma_j < 0$, its electron density will decrease to zero before or at flux $(p + 1)\phi_0$, so it has to merge with another subband. An example is the Chern number $-1$ subband at flux $2\phi_0$ in Fig. S14B, which merges with another subband at flux $3\phi_0$. If a subband at flux $p\phi_0$ has a Chern number $\sigma_j > 0$, it has to split into $\sigma_j + 1$ subbands ($\sigma_j > 0$) as flux increases by 1 (for example, The Chern number 2 subband at flux $2\phi_0$ and the Chern number 1 subbands at fluxes $3,4,5\phi_0$ in Fig. S14B). Therefore, a subband $j$ can adiabatically sustain itself as a single isolated subband for a wide range of magnetic fluxes only if its Chern number $\sigma_j = 0$ (e.g., the Chern number 0 subbands in Fig. S14B). Non-zero chern number bands have to split at some point as a function of the magnetic field.

These subbands carrying Chern number $\sigma_j = 0$, which remain a single isolated subband for a wide range of magnetic field, then give rise to the series of gaps $(4, s)$, $(8, s)$ ..., where $(C, s)$ denotes a gap with Chern number $C$ and electron filling $s = n(\phi = 0)/n_s$ at zero magnetic field, and $n_s = 4/\Omega_m$ with 4-fold spin-valley degeneracy considered. The reasoning is as follows:

First, given that the original LL gaps (in the untwisted limit) remain open in a wide range of sufficiently large magnetic field (large magnetic fluxes per moiré unit cell, but still much smaller than one magnetic flux per microscopic graphene unit cell), they correspond to $(4t, 0)$ gaps in the tBLG (counting the 4-fold spin-valley degeneracy). For example, gaps A and G in Fig. S14B correspond to the $(4,0)$ gap (occupying all the zero-mode LLs and below) and $(8,0)$ gap (occupying the zeroth and first LLs), respectively.

Then, in the illustrative Fig. S14B, the gaps C, D, E, F are separated with gap A by 1 to 4 Chern number 0 subbands, respectively. Accordingly, they correspond to a series of Chern number 4 gaps $(4,1)$, $(4,2)$, $(4,3)$, $(4,4)$. In particular, they can remain open over a wide range of magnetic fluxes, since the Chern number 0 subbands separating them can remain isolated without restriction from the Streda formula. In contrast, gap B (with quantum numbers $(0,3)$), which differs from gap A by a Chern number -1 band at flux $2\phi_0$, has to close readily at flux $3\phi_0$. This shows that the splitting of the first LL by moiré potential would most likely give rise to the $(4, s)$ series of gaps extending over a wide range of fluxes (and similarly $(8, s)$ for the second LL, and higher). For the zero-th LLs, similarly they could develop a Hofstadter spectrum with certain $(0, s)$ gaps remaining open in a wide range of magnetic fluxes at large enough magnetic field. Experimental data (main text Fig. 2) and numerical calculation (Fig. S12) show that this can happen, e.g., the $(0,0)$ and $(\pm 1,0)$ gaps (after they are interrupted by the $(\pm 4,0)$ gaps).

## 9. Tight-binding model for Fig. 3A-B

The Hofstadter butterfly in main text Fig. 3A-B are calculated with the following 2-band tight-binding model on a 2D square lattice (of lattice constant 1):

$$H_{\text{TB}}(\boldsymbol{k}) = \left(M - \cos k_x - \cos k_y\right)\sigma_z + A\left(\sigma_x \sin k_x + \sigma_y \sin k_y\right),$$

where $M$ and $A > 0$ are constants, $\sigma_{x,y,z}$ are Pauli matrices in the band basis, and $\boldsymbol{k} = (k_x, k_y)$ is the quasimomentum in the BZ. We note that this model can be viewed as half of the Bernevig-Hughes-Zhang model (33).

The valence band and conduction band of this model both have Chern numbers 0 if $|M| > 2$. In contrast, when $|M| < 2$, the valence band and conduction band have Chern numbers $\pm M/|M|$, respectively. Fig. 3A is calculated by setting $M = 3$ and $A = 2$, for which both bands are trivial bands with Chern number 0. Fig. 3B is calculated by setting $M = 1$ and $A = 2$, for which the valence and conduction bands carry Chern number $\pm 1$, respectively.

**Fig. S1.**

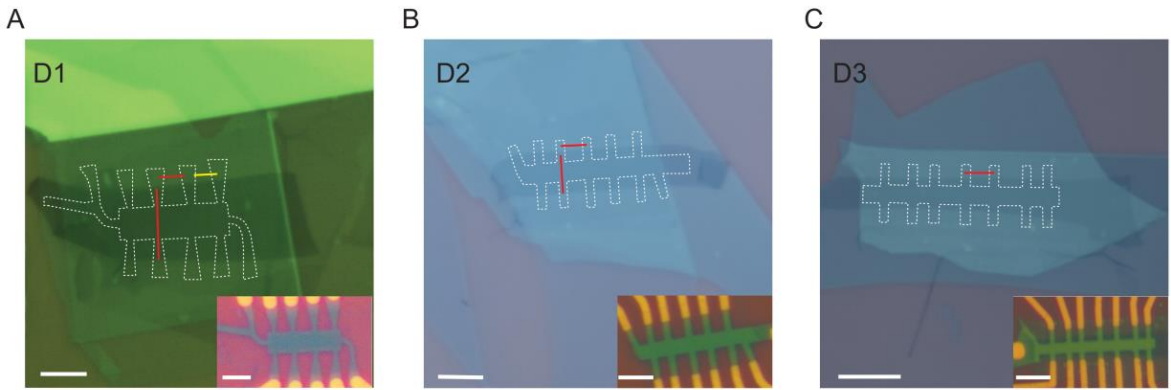

Fig. S1. **Optical microscope images of stacks and devices.** Red lines in (**A**) mark the contact pairs which are further labeled as device D1-1 (0.45°). The contact pair indicated by yellow line is labeled as D1-2 (0.44°) with measured data shown in Fig. S8B. Device D2 (0.38°) and device D3 (0.34°) are measured with contact pairs indicated by red lines in (**B**) and (**C**), with corresponding results displayed in Fig. S7 and S8. The outlines of fabricated Hall bars are demonstrated with the white dashed lines. All scale bars are 5μm.

**Fig. S2.**

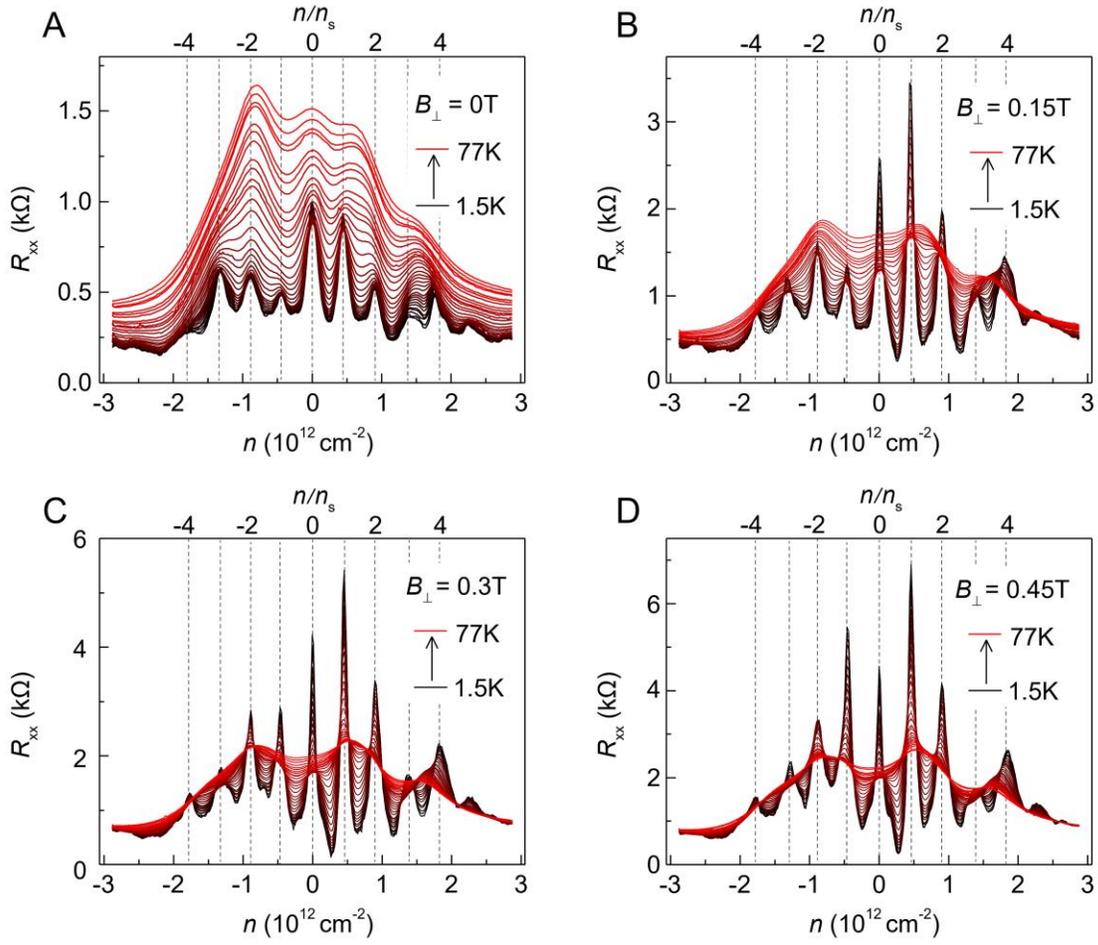

Fig. S2. **Magnetic field induced metal-insulator transition.** Longitudinal resistance $R_{xx}$ vs. carrier density $n$ at various temperatures measured at a magnetic field $B_\perp$ of 0T (**A**), 0.15T (**B**), 0.3T (**C**) and 0.45T (**D**). Data is measured from D1-1 (0.45°).

**Fig. S3.**

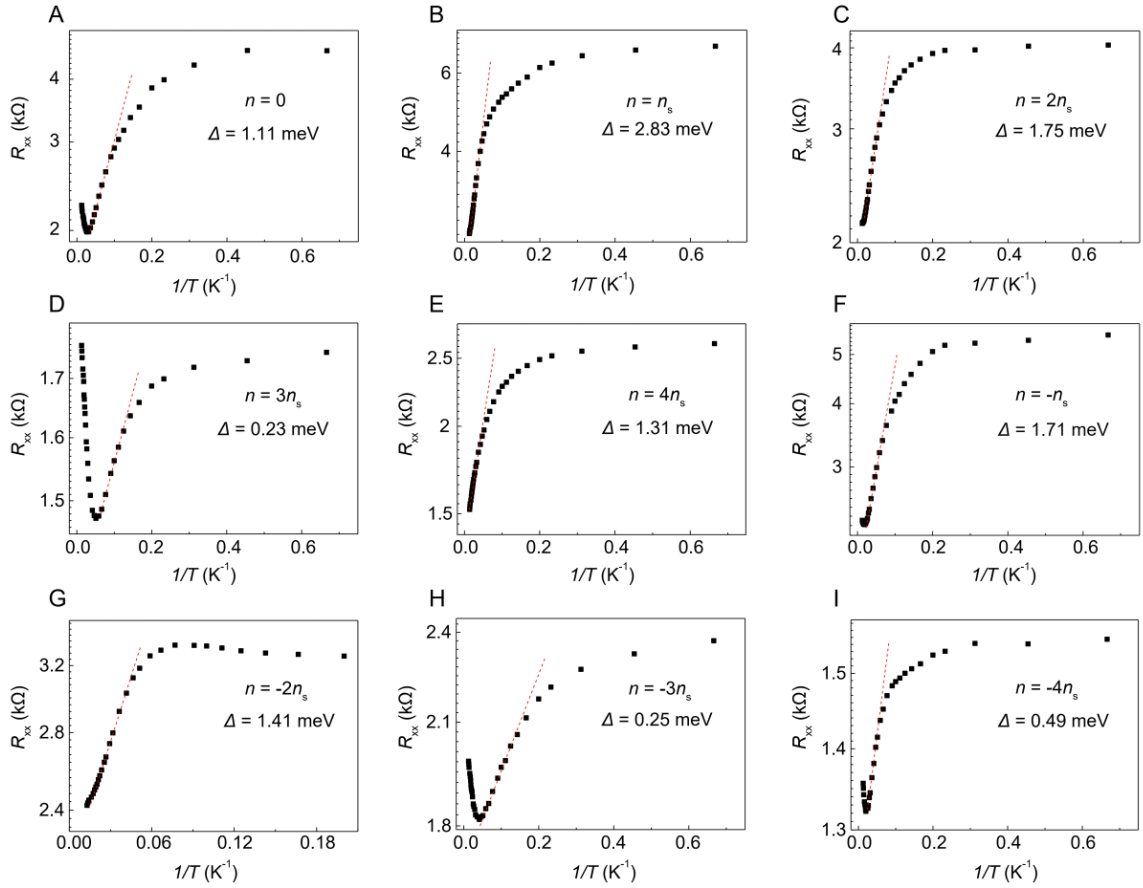

Fig. S3. **Extraction of gaps in all integer filling states.** (**A-I**) Longitudinal resistance $R_{xx}$ vs. inverse temperature $1/T$ at various integer fillings $n=Nn_s$ and $B_\perp = 0.45$T. The straight dashed lines are fits to $R_{xx} \sim \exp(\Delta/2kT)$ temperature activated behavior. Data is measured from D1-1 (0.45°).

**Fig. S4.**

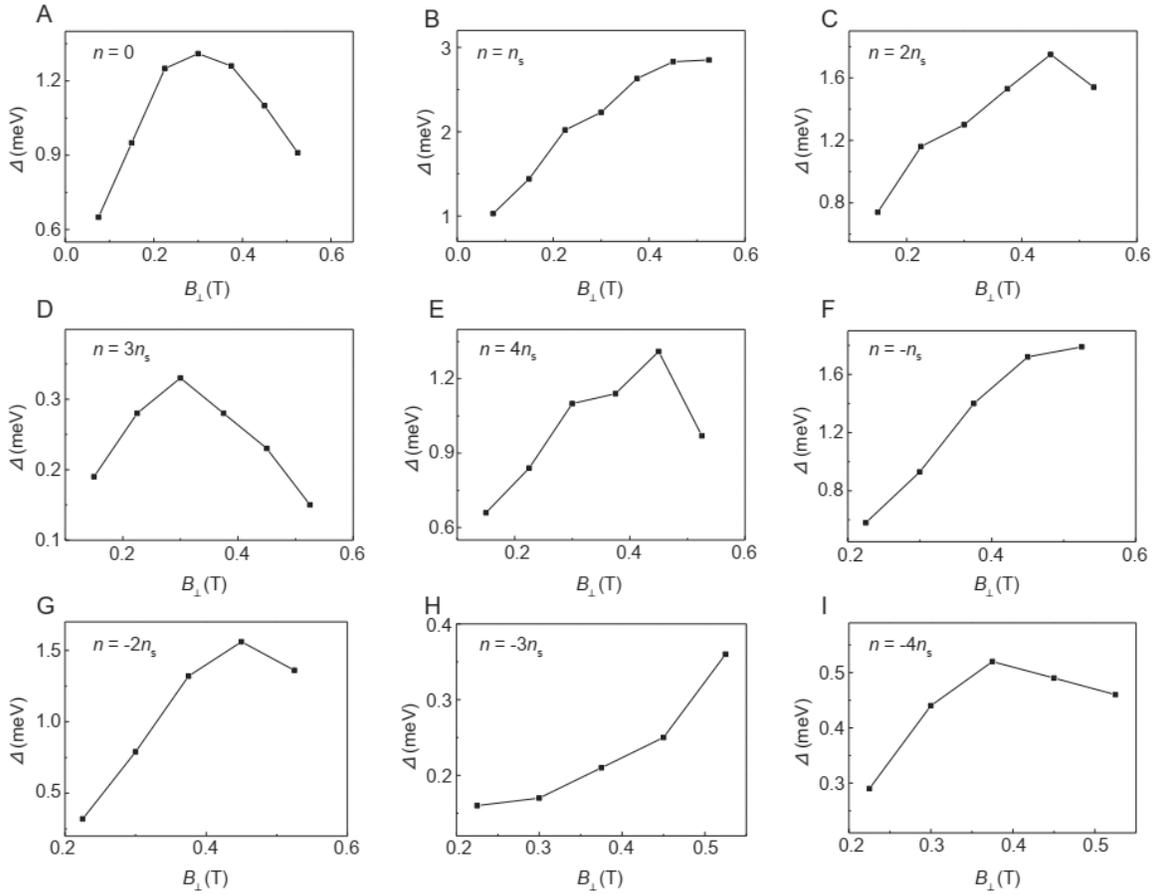

Fig. S4. **Gap evolution with $B_\perp$ at all integer filling stages.** (A-I) Magnetic field $B_\perp$ dependent gaps at various integer fillings $n=Nn_s$ extracted from $R_{xx} \sim \exp(\Delta/2kT)$ temperature activated behavior. Data is measured from D1-1 (0.45°).

**Fig. S5.**

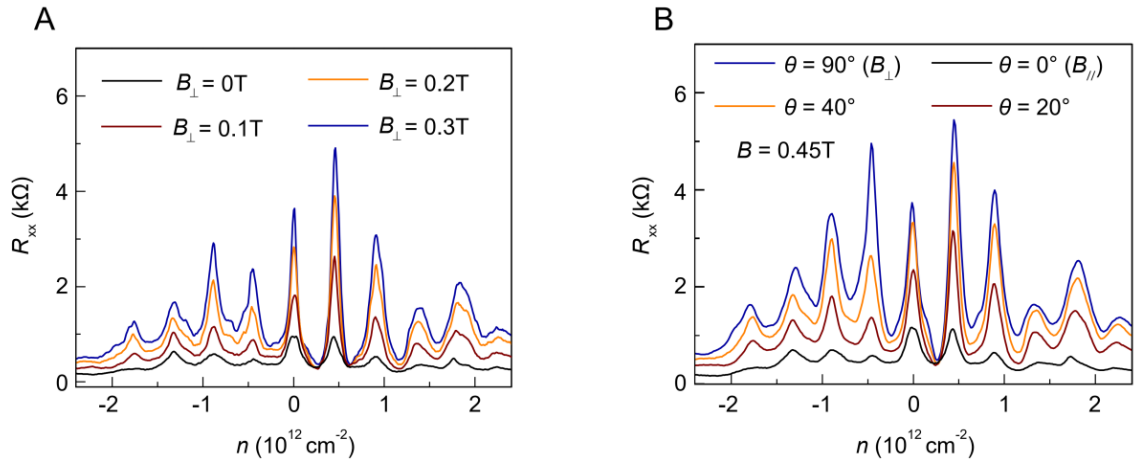

Fig. S5. **$R_{xx}$ vs. carrier density $n$ at various magnetic field.** (**A**) $R_{xx}$ vs. $n$ at different perpendicular magnetic field $B_\perp$. (**B**) $R_{xx}$ vs. $n$ at fixed total field $B = 0.45$T with different tilt angles. Data is measured from D1-1 (0.45°) at a temperature of 1.5K

**Fig. S6.**

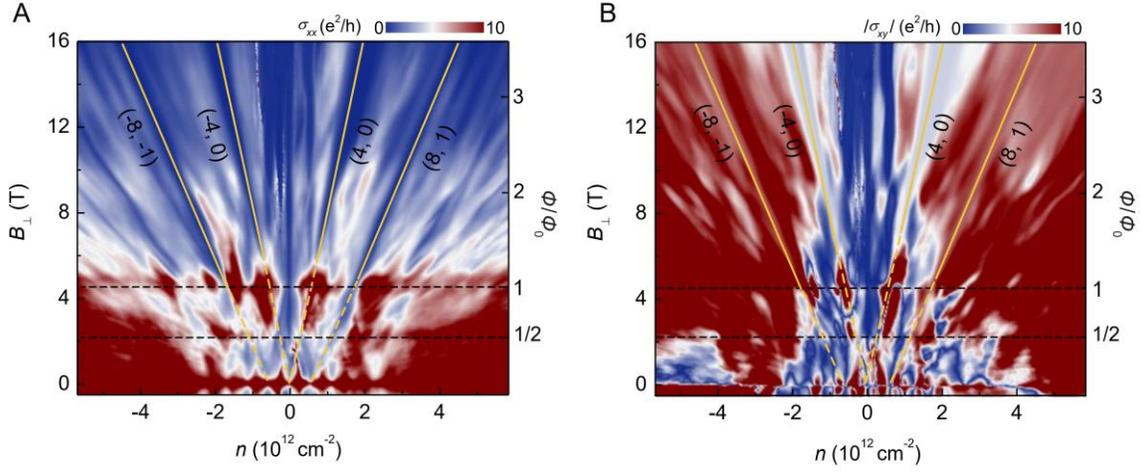

Fig. S6. **Hofstadter butterfly measured from D1-1.** Color plot of longitudinal conductivity $\sigma_{xx} = \rho_{xx}/(\rho_{xx}^2 + R_{xy}^2)$ (**A**) and magnitude of hall conductivity $/\sigma_{xy}/ = /R_{xy}//(\rho_{xx}^2 + R_{xy}^2)$ (**B**) as a function of carrier density $n$ and $B_\perp$ measured at a temperature of 1.5K. Yellow solid lines indicate gaps with different Chern numbers and yellow dashed lines give sight guidance for tracing to different moiré bands. Black dashed lines mark the positions of magnetic flux values ($\phi_0/2$ and $\phi_0$) through the moiré unit cell.

**Fig. S7.**

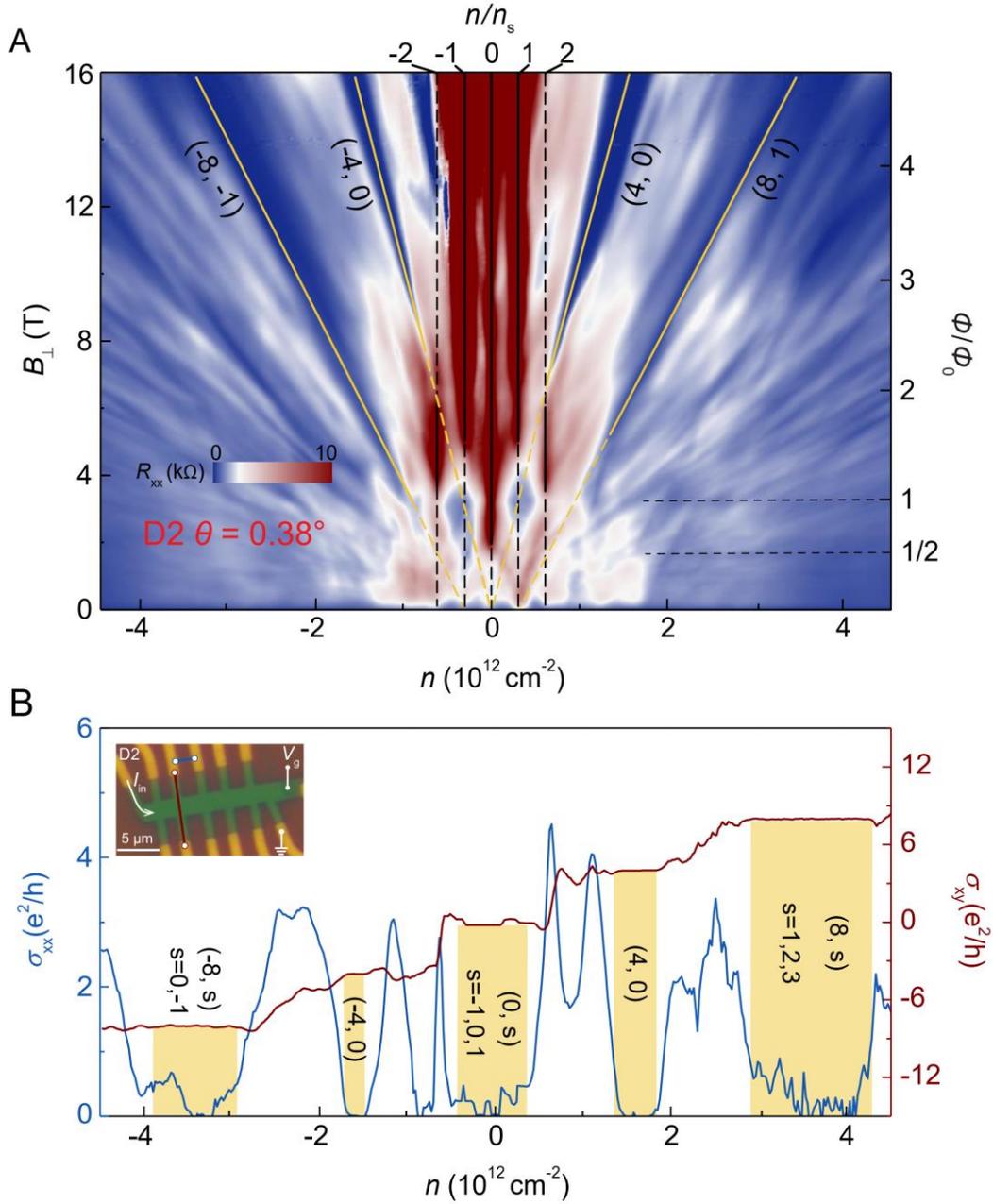

Fig. S7. **Hofstadter butterfly spectrum in 0.38° tBLG (D2).** (**A**) Color plot of longitudinal resistance $R_{xx}$ as a function of carrier density $n$ and $B_\perp$. Black and yellow solid lines indicate gaps with different Chern numbers and extended dashed lines give sight guidance for tracing to different moiré bands. The horizontal dashed lines mark the positions of magnetic flux values ($\phi_0/2$ and $\phi_0$) through the moiré unit cell. (**B**) Longitudinal conductivity $\sigma_{xx}$ and hall conductivity $\sigma_{xy}$ vs. carrier density $n$ measured at a maximum magnetic field of 16T. Both (**A**) and (**B**) are measured at a temperature of 1.5K.

**Fig. S8.**

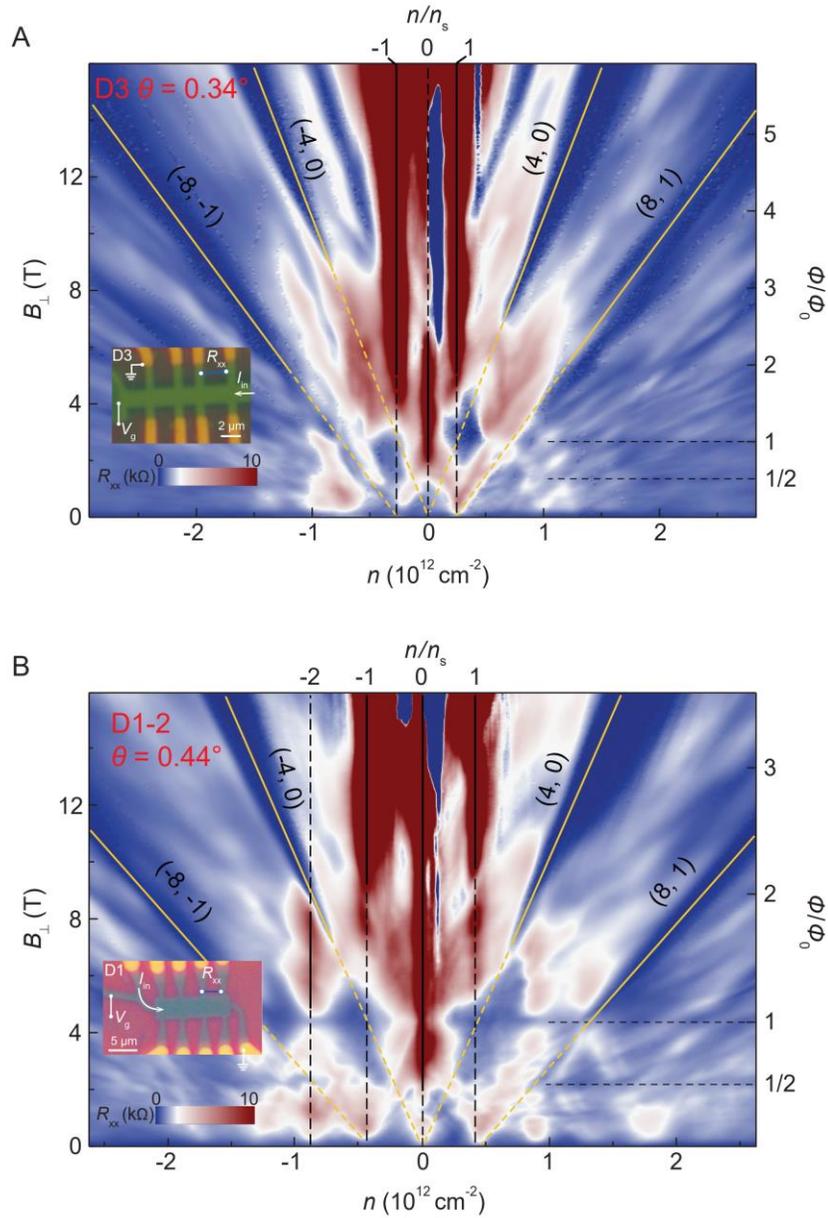

Fig. S8. **Hofstadter butterfly spectrums measured from device D3 (0.34°) and D1-2 (0.44°).**
Color plot of longitudinal resistance $R_{xx}$ as a function of carrier density $n$ and $B_\perp$ measured from device D3 (**A**) and device D1-2 (**B**). Inserts show optical images and measurement configurations of devices D3 and D1-2. Black and yellow solid lines indicate gaps with different Chern numbers and extended dashed lines give sight guidance for tracing to different moiré bands. The horizontal dashed lines mark the positions of magnetic flux values ($\phi_0/2$ and $\phi_0$) through the moiré unit cell. Both (**A**) and (**B**) are measured at a temperature of 1.5K.

**Fig. S9.**

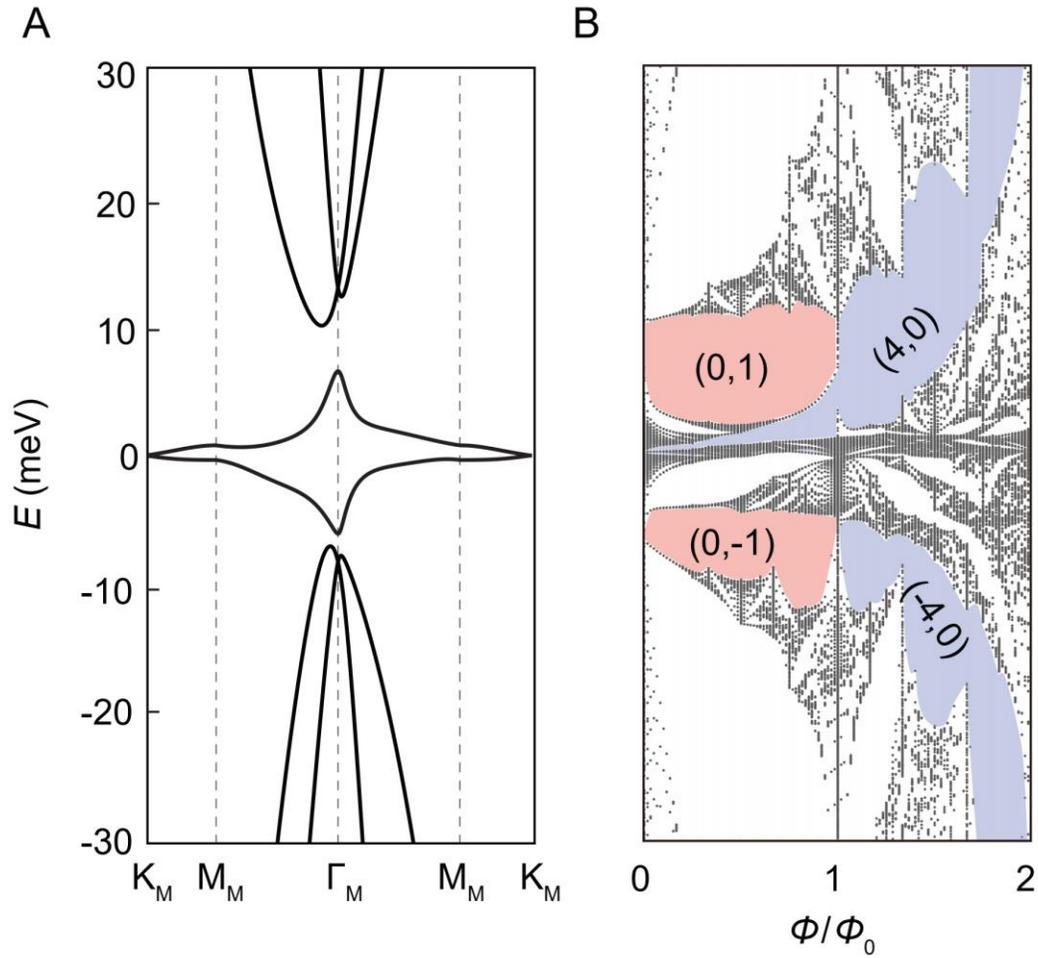

Fig. S9. (**A**) The band structure of the ten-band tight-binding model (one spin one valley) in Ref (*8*) at zero magnetic field. (**B**) The Hofstadter butterfly of the ten-band tight-binding model, where $(C, s)$ labels a Hofstadter gap with Chern number $C$ (counting the 4-fold spin-valley degeneracy) and zero-magnetic-field filling $s = n/n_s$ per unit cell. Both figures are replotted from Ref (*14*).

**Fig. S10.**

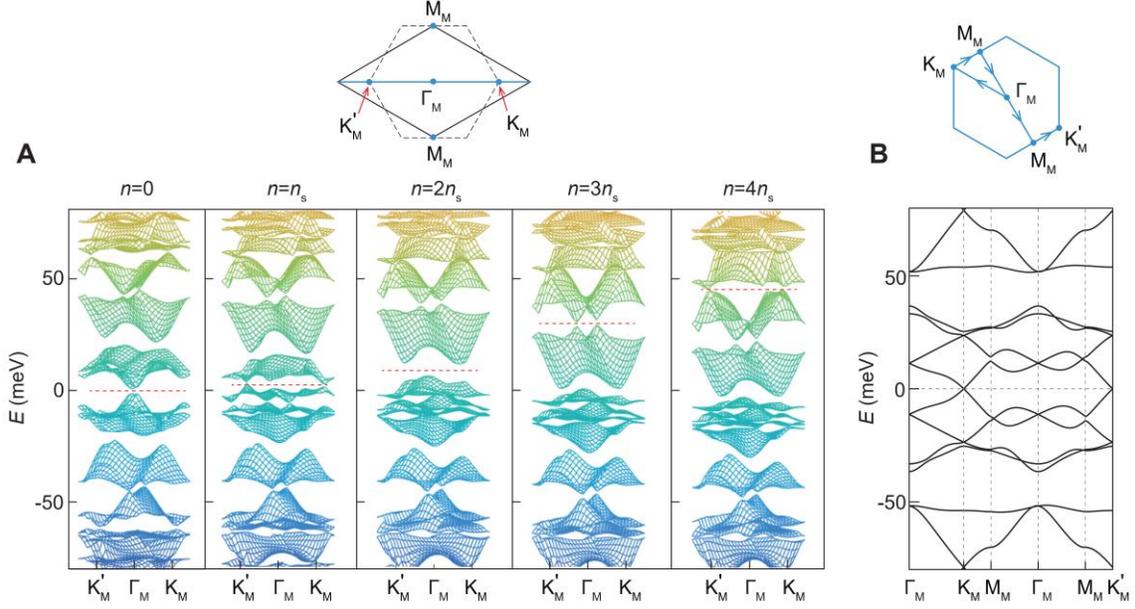

Fig. S10. **Band structures simulation**. (**A**) The band structure of the 0.45° tBLG under the Hartree-Fock approximation considering Coulomb interaction $\frac{|g_1|V(g_1)}{\sqrt{3}} = 10$ meV between electrons ($g_1$: moiré reciprocal vector) at various integer fillings $n$=N$n_s$, where the Fermi energies are indicated with red dashed lines. The 2D band structures are plotted in the diamond (instead of honeycomb) moiré Brillouin zone (see the top insert in (**A**)), and viewed from the side perpendicular to the long diagonal of the diamond. For each filling, the $C_{3z}$ symmetry is spontaneously broken, and $C_{2z}T$ is preserved, allowing Dirac points away from high symmetry points of the moiré Brillouin zone. The first panel for $n = 0$ is also shown in the main text Fig. 1(**C**). (**B**) The non-interacting band structure of tBLG calculated using the continuum model (without interaction), and plotted along the moiré Brillouin zone high symmetry lines shown in the top insert of (**B**). The lowest 8 bands are connected among each other by Dirac points, and are gapped from higher bands, which is different from the Hartree-Fock interacting band structures in (**A**).

**Fig. S11.**

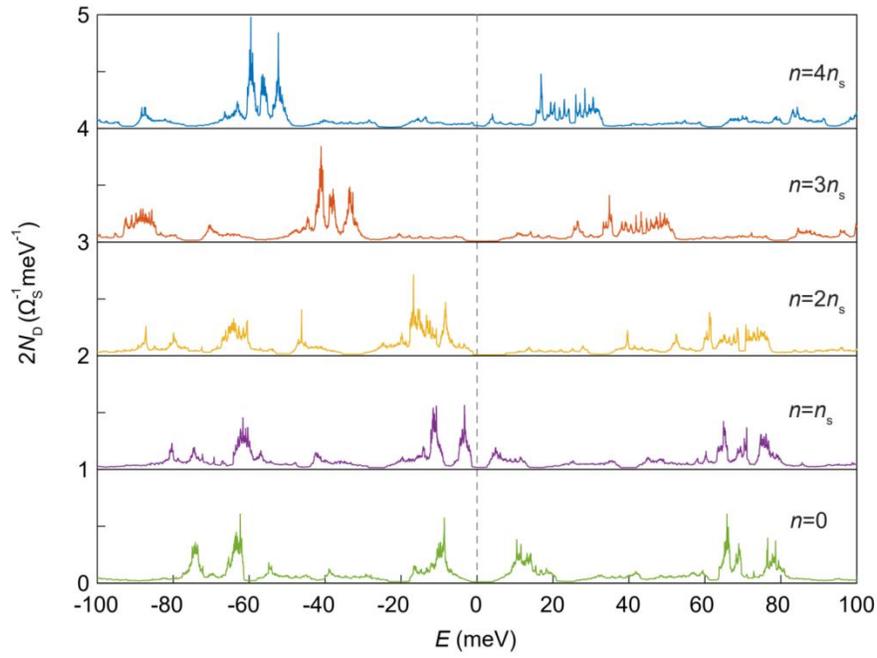

Fig. S11. Density of states in 0.45° tBLG continuum model with Coulomb interactions between electrons considered using the Hartree-Fock method at zero magnetic field, where we assume no spin or valley degeneracy breaking. The calculations are done for full moiré band fillings $n = Nn_s$. The corresponding Hartree-Fock band structures are shown in Fig. S10A.

## Fig. S12.

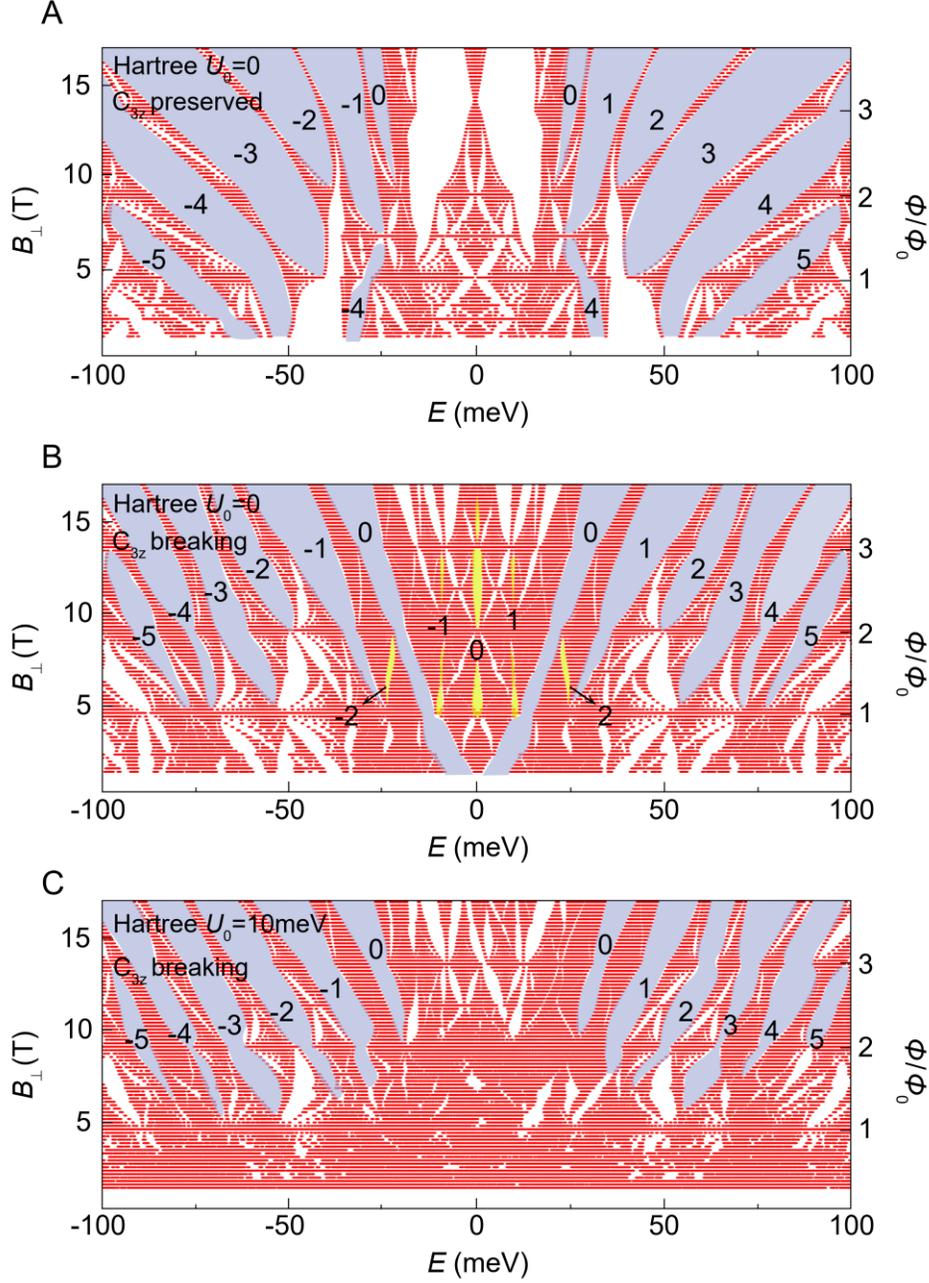

Fig. S12. **Calculated Hofstadter Butterfly of 0.45° tBLG.** Hofstadter butterfly of the tBLG continuum model (one-spin one-valley) under different Hartree potential and $C_{3z}$ configurations. In (**B**) and (**C**), a 0.4% strain is considered and $C_{3z}$ symmetry is broken by a single particle term. A 10meV Hartree potential is further considered in (**C**). $C = \pm 4$ gaps are highlighted with light blue color with corresponding numbers denoting different $s$ values. Yellow color in (**B**) highlights $C=0$ gaps ($s = 0, \pm1, \pm2$) which reappear at high $B_\perp$ field ($\phi > \phi_0$).

**Fig. S13.**

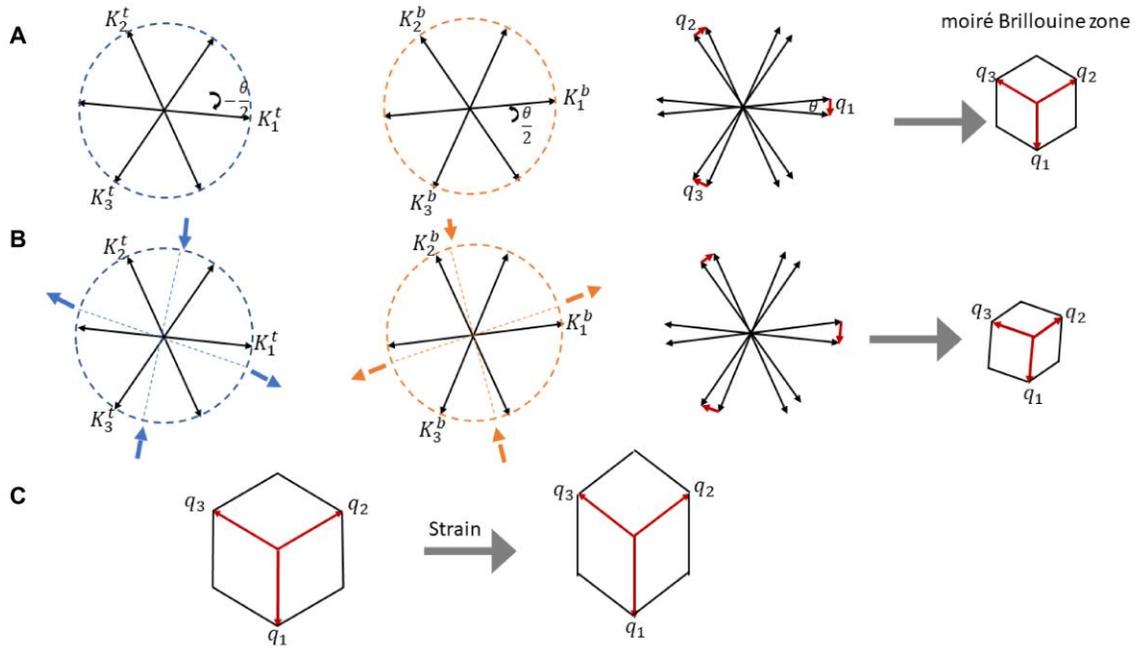

Fig. S13. **Strained tBLG.** (**A**) Momentum space of tBLG without strain. The momentum space of top layer of graphene rotated rotated by $-\frac{\theta}{2}$ and top layer rotated by $\frac{\theta}{2}$. The momentum difference of the Dirac points of the two layers is used to construct the moiré Brillouin zone. (**B**) The rotated top and bottom layers are under independent uniaxial strain. The resultant momentum difference between the Dirac points are no longer related by $C_{3z}$ rotation symmetry. (**C**) The specific choice of strained moiré Brillouin zone used in the Hofstadter butterfly calculations.

**Fig. S14.**

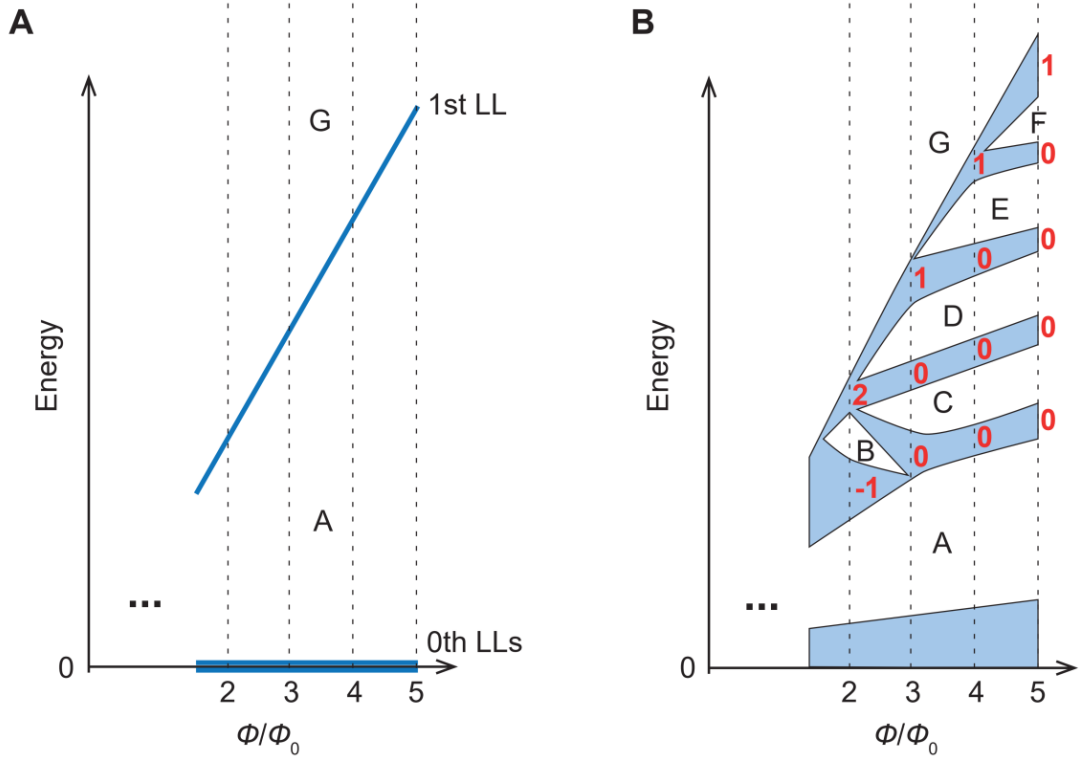

Fig. S14. (**A**) Illustration of the LLs in the untwisted bilayer graphene at nonzero magnetic field. Counting the 4-fold spin-valley degeneracy, the LL gap A carries Chern number $C = 4$, and the LL gap G has Chern number $C = 8$. At zero magnetic field, all the LL gaps collapse to the CNP. (**B**) Illustration of large magnetic field Hofstadter butterfly of twisted bilayer graphene, which can be viewed as the LLs of the untwisted bilayer graphene splitting into moiré subbands, during which we assume the LL gaps of untwisted bilayer graphene (e.g., A and G) remain open (above certain threshold magnetic field). $\phi/\phi_0$ is the number of magnetic fluxes per moiré unit cell. For illustration purpose, only the splitting of the 1st LL is shown, while the splitting of the 0th LLs (per spin per valley) are not shown. The red numbers near the dashed vertical lines denote the Chern number $\sigma_j$ of the subbands (per spin per valley) at integer magnetic fluxes (the dashed vertical lines).